# A physics-regularized neural network and Kirchhoff Markov random field framework for inferring latent electrochemical states from operando spectromicroscopy

Naoki Wada[1]†, Yuta Kimura[2], Masaichiro Mizumaki[3], Koji Amezawa[2], Ichiro Akai[4] and Toru Aonishi[1]†*

[1]Department of Complexity Science and Engineering, Graduate School of Frontier Sciences, The University of Tokyo, Kashiwa, Chiba 277-8561, Japan

[2]Institute of Multidisciplinary Research for Advanced Materials, Tohoku University, Sendai, Miyagi 980-8577, Japan

[3]Graduate School of Science and Technology, Kumamoto University, Kumamoto, Kumamoto 860-8555, Japan

[4]Institute of Industrial Nanomaterials, Kumamoto University, Kumamoto, Kumamoto 860-8555, Japan

†These authors contributed equally to this work.
*Correspondence: aonishi@edu.k.u-tokyo.ac.jp

**Abstract**

Quantitative understanding of coupled reaction-transport dynamics in lithium-ion battery (LIB) composite electrodes is limited by the inaccessibility of key internal electrochemical states. Here we present a physics-integrated, data-driven framework to reconstruct latent states from operando microscopic X-ray absorption fine structure ($\mu$-XAFS) hyperspectral data of LIB cathodes. Normalized-local-lithium-stoichiometry (NLLS) maps derived from Co K-edge spectra are refined using a physics-regularized neural network that enforces spatial continuity and current conservation to resolve ambiguities in the weak-sensitivity region. The reconstructed NLLS dynamics are embedded in a Kirchhoff-based Markov random field incorporating Kirchhoff's laws, Ohm's law, and Butler-Volmer kinetics to infer interfacial current density, ionic current, electrolyte potential, and effective ionic conductivity. Application to electrodes with differential initial electrolyte concentrations reveals distinct reaction-propagation modes that are consistent with a mechanism involving the non-monotonic concentration dependence of electrolyte conductivity. The inferred electrolyte concentration profiles qualitatively resemble the spatially extended concentration changes observed by independent operando X-ray transmission imaging using a 1 M $LiAsF_6$ electrolyte.

# 1. Introduction

In response to the increasing performance demands of portable electronic devices, electric micromobility systems, and drones, as well as the widespread adoption of electric vehicles and the large-scale integration of renewable energy sources, there is a strong demand for the development of lithium-ion rechargeable batteries (LIBs) with high energy density and high reliability [1-6]. To meet these requirements, it is essential to understand in detail the mechanisms of charge-discharge reactions occurring within LIB electrodes and to optimize electrode structures and material design based on the knowledge obtained.

The charge-discharge reactions in LIB composite electrodes are highly complex processes involving strongly coupled phenomena, including lithium-ion transport in the electrolyte, electron transport through conductive additives and active materials, charge transfer at the active material/electrolyte interface, and lithium diffusion within active material particles [7]. Moreover, owing to the complex microstructures formed by active materials, conductive additives, and binders, as well as the intricate ionic and electronic conduction pathways, these reactions proceed spatially inhomogeneously within the electrode. Their spatial distributions evolve dynamically as charging and discharging progress [8-11].

Against this background, attempts have been made to directly capture the overall behavior of charge-discharge reactions in composite electrodes using advanced characterization techniques. For example, spatial inhomogeneity of lithium concentration (Li heterogeneity) in active materials has been evaluated using scanning

/ full-field synchrotron X-ray microspectroscopy [12–16]. In addition, spatially inhomogeneous electrolyte salt concentration distributions arising from local stagnation of lithium-ion transport in the electrolyte have been visualized using X-ray fluorescence analysis [17], magnetic resonance imaging [18,19], and phase-contrast X-ray imaging [20]. However, these techniques capture only one aspect of electrode reactions and have not led to a comprehensive understanding of the entire reaction process in which multiple transport and reaction processes are strongly coupled.

On the other hand, numerical simulations based on porous electrode models have been developed to reproduce charge–discharge reactions within electrodes and to infer the underlying mechanisms [21–27]. In particular, lithium-ion transport simulations using electrode pore structures obtained by X-ray CT [23], simulations incorporating the structure of the carbon binder domain phase obtained by X-ray CT [24,25], and the development of Python libraries for implementing such simulations [27] have been reported. While these numerical simulations are valuable, important parameters such as exchange current density are often not sufficiently determined experimentally and must be assumed. Moreover, the validity of simulation results has typically been assessed only indirectly through their consistency with macroscopic electrode responses, such as charge–discharge curves and capacity. Direct verification of the spatiotemporal dynamics of reactions occurring inside the electrode remains difficult.

In recent years, data-driven scientific approaches based on machine learning have been increasingly employed to extract latent structures embedded in hyperspectral images acquired by techniques such as scanning transmission electron microscopy–electron energy loss spectroscopy (STEM-EELS), Raman spectroscopy, and synchrotron microspectroscopy [28–34]. Applications include compound labeling and noise reduction using non-negative matrix factorization (NMF) [28–30,33,34], analysis of electronic state distributions by combining Bayesian spectroscopy [34–36] with NMF [31], and phase behavior analysis in conjunction with operando pair distribution function measurements [32]. Applying such methods to spatiotemporal measurement data of charge–discharge reactions may enable simultaneous extraction of insights into the material transport and charge-transfer processes governing the reactions.

In this study, we develop an analysis pipeline to reconstruct hidden electrochemical states in lithium-ion battery (LIB) composite cathodes from operando microscopic X-ray absorption fine structure (μ-XAFS) hyperspectral data. By leveraging the spatiotemporal spectral evolution associated with charge–discharge reactions, the framework enables quantitative inference of internal transport variables that are not directly accessible from spectroscopy alone, including solid-liquid interfacial current density and potential difference, liquid-phase ohmic resistance, and electrolyte concentration dynamics.

The pipeline consists of two components. First, a physics-regularized three-layer neural network (NN) estimates normalized-local-lithium-stoichiometry (NLLS) distributions, including regions with minimal spectral sensitivity to lithium insertion/extraction. Spatial continuity of NLLS and current conservation are imposed as constraints to ensure stable reconstruction. Second, the inferred NLLS dynamics are incorporated into a Kirchhoff-Markov random field (MRF) framework based on a simplified porous electrode equivalent circuit. By integrating Kirchhoff's laws, Ohm's law, and interfacial reaction kinetics within a probabilistic formulation, the model consistently estimates internal transport-related state variables.

The method is applied to operando datasets obtained from composite electrodes with three different initial $LiPF_6$ electrolyte concentrations [15], enabling reconstruction of the spatiotemporal evolution of transport states and quantitative examination of concentration-dependent reaction heterogeneity. Independent operando X-ray transmission imaging using $LiAsF_6$ electrolyte, in which concentration can be directly quantified from transmission contrast, shows qualitative agreement with the reconstructed concentration profiles, providing an independent qualitative comparison.

This paper is organized as follows. Section 2 describes the experimental procedures and analytical methods. Section 3 presents the reconstructed internal states and their independent qualitative consistency check. Section 4 discusses the mechanism underlying concentration-dependent spatial charge distribution [15].

# 2. Methods

## 2.1 Operando $\mu$-XAFS Measurements of LIB Composite Cathodes Using $LiPF_6$ Electrolyte

In this study, we analyzed operando $\mu$-XAFS data acquired from the model composite cathode reported by Nakamura et al. [15]. The model electrode had a planar area of 1 $cm^2$ and a thickness of 46 μm and consisted of $LiCoO_2$ (LCO), acetylene black (AB), and a binder at a mass ratio of 80:10:10. It was sandwiched between an aluminum current collector and a polyimide insulating sheet. The electrolyte was $LiPF_6$ dissolved in a mixture of ethylene carbonate (EC) and dimethyl carbonate (DMC), and lithium metal foil was used as the negative electrode.

This electrode configuration permits electron transport throughout the electrode while restricting ionic exchange with the surrounding electrolyte to the four lateral edges. Consequently, the charge-discharge reaction propagates from the electrode edge toward the interior and can locally be regarded as a one-dimensional process (Fig. 1(a1)).

The cell was charged at a constant current of 110 μA until the voltage reached the cutoff value of 4.3 V. Based on the theoretical specific capacity of LCO (274 mAh $g^{-1}$), the LCO density (5.16 g $cm^{-3}$), and the LCO volume fraction in the electrode (0.284), the theoretical capacity of the electrode was estimated to be 1.846 mAh. The applied current therefore corresponds to a nominal whole-electrode charging rate of approximately 0.060 C. Normalizing the applied current by the total cross-sectional area of the four lateral edges, $4 \times 1\,\mathrm{cm} \times 46\,\mu\mathrm{m}$, gives an edge-area-normalized current density of approximately 6 mA $cm^{-2}$. This value characterizes the lateral ionic-current geometry and should be distinguished from the conventional C-rate based on the total electrode capacity. Because substantial delithiation occurred only within part of the electrode, the nominal whole-electrode C-rate does not necessarily represent the effective local charging rate in the electrochemically active region.

Nakamura et al. visualized the one-dimensional propagation of the charging reaction by irradiating X-rays along the electrode-thickness direction, perpendicular to the direction of reaction propagation. Co K-edge XAFS hyperspectral images were acquired at each time point during charging (Fig. 1(a2)). To enable operando measurements, the energy range was restricted to the vicinity of the white line at 7727–7732 eV (Fig. 1(b1)). The observation area near the electrode edge was 600 × 1200 $\mu m^2$. The spatial sampling interval was 6.5 μm, and the temporal sampling interval was 300 s for the initial electrolyte concentration of 0.3 M and 240 s for the 1 M and 2 M conditions. Further details of the measurements and sample preparation are provided in Ref. [15].

## 2.2 X-ray Transmission Imaging of LIB Composite Cathodes Employing $LiAsF_6$ Electrolyte

Operando X-ray transmission imaging was performed using a model composite cathode with the same basic edge-fed geometry as that employed in the $\mu$-XAFS measurements described in Section 2.1. The electrode had a planar area of 1 $cm^2$ and a thickness of approximately 50 μm. It was fabricated from $LiCoO_2$, acetylene black, and poly(vinylidene fluoride) (PVDF) at a mass ratio of 80:10:10. The resulting cathode slurry was coated onto an aluminum current collector, covered with a polyimide insulating film on the opposite surface, and dried. This configuration restricted ionic exchange with the surrounding electrolyte primarily to the four lateral edges of the electrode.

Laminate cells were constructed using the model composite cathode, lithium metal as the negative electrode, and an electrolyte consisting of 1 mol $L^{-1}$ $LiAsF_6$ dissolved in ethylene carbonate-ethyl methyl carbonate (EC-EMC, 3:7 v/v). The cells were charged for 45 min at a constant current of 120 μA. Assuming equal current distribution among the four lateral edges, normalization by their total cross-sectional area, $4 \times 1\,\mathrm{cm} \times 50\,\mu\mathrm{m}$, gives an edge-area-normalized current density of approximately 6 mA $cm^{-2}$.

X-ray transmission images of the composite cathode were acquired at 1, 5, 25, 35, and 45 min after the start of charging using 6.5 keV X-rays with a beam size of approximately 220 × 800 μm. The spatial resolution of the operando X-ray transmission imaging was 1.3 μm. Because arsenic has a relatively high X-ray absorption coefficient, changes in the $LiAsF_6$ electrolyte concentration produce measurable changes in the transmission-image contrast. Spatial changes associated with the electrolyte concentration were therefore evaluated from the difference in image intensity between each charging time point and the corresponding image acquired before charging in the same observation region.

## 2.3 NLLS Evaluation Based on PTE and the Weak-Sensitivity Region

In the previous study by Nakamura et al. [15], NLLS maps were constructed from XAFS hyperspectral images using the procedure described below. The same procedure was adopted for the initial NLLS evaluation in the present study.

Figure 1(b1) shows a representative Co K-edge XAFS spectrum acquired at a single spatiotemporal point. In LCO, lithium extraction is generally accompanied by an increase in the Co valence state and a corresponding shift of the peak-top energy (PTE) of the Co K-edge toward higher energy. The PTE was therefore extracted from the XAFS spectrum at each spatiotemporal point, and the corresponding normalized local lithium stoichiometry (NLLS), $x$, was evaluated using the quadratic regression curve obtained from standard samples, as shown in Fig. 1(b2).

However, above the threshold $x_{th}$, the PTE changes only slightly with further lithium extraction, making it difficult to determine NLLS uniquely from the PTE. This weak-sensitivity range overlaps with the two-phase reaction region of LCO reported in previous studies [37,38]. We therefore refer to this NLLS interval operationally as the weak-sensitivity region. For $x \geq x_{\mathrm{th}}$, NLLS cannot be determined uniquely from the regression curve and is provisionally assigned a representative value of 1, whereas for $x < x_{\mathrm{th}}$, NLLS is evaluated directly from the regression curve. This procedure produces an NLLS map at each measurement time, as exemplified in Fig. 1(b3). This assigned value serves only as an initial provisional estimate and is subsequently refined using the physics-regularized neural network described in the following section.

## 2.4 Overview of the Analysis Pipeline

In this study, we constructed an analysis pipeline to quantitatively extract material transport and charge-transfer processes coupled with charge-discharge reactions from XAFS hyperspectral data obtained by operando microspectroscopy of LIB composite cathodes. As a proof-of-concept model case, we analyzed Co K-edge XAFS spectral data of the above-mentioned LIB composite cathodes acquired by operando $\mu$-XAFS measurements [15].

Figure 1(c) presents the overall workflow of the proposed pipeline. The pipeline takes as input the NLLS maps constructed using the conventional method by Nakamura et al. The pipeline comprises two main components:

(i) a three-layer NN that resolves ambiguities in the weak-sensitivity region where

NLLS cannot be uniquely determined, and
(ii) a Kirchhoff-based Markov random field (Kirchhoff-MRF) that quantitatively estimates material transport processes based on the inferred NLLS distribution.

In the NN module, physics-informed regularization is introduced to enforce spatial continuity of NLLS and current conservation, enabling stable reconstruction of the spatiotemporal NLLS distribution, including the weak-sensitivity region. The inferred NLLS dynamics are then supplied to the Kirchhoff-MRF, which integrates Kirchhoff's current and voltage laws, Ohm's law, and interfacial reaction kinetics within a simplified porous electrode model. This probabilistic framework allows consistent estimation of internal physical quantities constrained by both electrochemical law and the NLLS distribution.

This integrated approach enables quantitative inference of otherwise inaccessible internal state variables, including solid-liquid interfacial current density, ionic current in the electrolyte, electrolyte potential distribution, and effective ionic conductivity.

Detailed descriptions of each component are provided in the following sections.

## 2.5 Physics-regularized NN for NLLS estimation in the Weak-Sensitivity Region

With the method described in Section 2.3, NLLS in the weak-sensitivity region cannot be uniquely determined, which limits quantitative analysis of lithium-ion transport processes. To overcome such difficulties, we introduce spatial continuity of NLLS and current conservation as regularization constraints, and regress the spatiotemporal NLLS distribution using a three-layer NN under physics-informed regularization.

The data structure of the NLLS spatiotemporal map is shown in Fig. 2(a). We define the NLLS at time $t$, at pixel index $i$ in the $z$-direction corresponding to the direction from the electrode edge toward the center, i.e., the direction in which charging proceeds, and at pixel index $j$ in the lateral direction as $x_{i,j}(t)$. Let the numbers of pixels in the $z$- and lateral directions be $N$ and $M$, respectively, and the total number of frames be $L$. Because the reaction progression considered in this study is one-dimensional, we treat the lateral pixel index $j$ as an ensemble sample index, and focus on the $z$-direction pixel index $i$ and time $t$ in the following.

The structure of the introduced three-layer NN is shown in Figs. 2(b1) and 2(b2). In Fig. 2(b1), the input layer consists of cells $In_i$ corresponding to each pixel index $i$ in the $z$-direction, and outputs time $t$. The hidden layer has $K$ cells $M_{1,i}$ to $M_{K,i}$ for each pixel index $i$, and uses a sigmoid activation function. The output layer consists of cells $O_i$ corresponding to each pixel index $i$, and outputs the estimated NLLS $y_{O,i}(t)$ without using an activation function. Here, ① is a cell that outputs the constant 1 and corresponds to the bias term. Accordingly, the input-output relationship of the NN for pixel index $i$ is given by Eq. (1):

$$y_{O,i}(t) = \sum_{k=1}^{K} b_{k,i}\, \sigma\big(a_{k,1,i} - a_{k,2,i}\, t\big) + b_{K+1,i}. \qquad (1)$$

Here, $\sigma(x)$ is the sigmoid function defined as $\sigma(x) = (1 + \tanh(x))/2$. The parameters $\{a_{1,1,i}, a_{1,2,i}\}$ to $\{a_{K,1,i}, a_{K,2,i}\}$ represent the connection weights to the cells $M_{1,i}$ to $M_{K,i}$, and $\{b_{1,i}, \ldots, b_{K,i}, b_{K+1,i}\}$ are the connection weights to the cell $O_i$. In this study, because many reaction curves can be described by hyperbolic functions, we adopted the sigmoid function as the activation function.

The time derivative of the NN output $y_{O,i}(t)$ corresponds to the current associated with lithium deintercalation/intercalation in LCO at pixel index $i$, and is expressed by Eq. (2):

$$I_{O,i}(t) = -\frac{S}{2} \sum_{k=1}^{K} \frac{a_{k,2,i}\, b_{k,i}}{\cosh^2\big(a_{k,1,i} - a_{k,2,i}\, t\big)}. \qquad (2)$$

Here, $S$ converts the time derivative of the estimated NLLS into the LCO current density $I_{O,i}(t)$. The current associated with each spatial subregion is normalized by the electrode cross-sectional area perpendicular to the $z$-direction, $A_{prep} = wh$, where $w$ is the length of the modeled electrode edge and $h$ is the electrode thickness. Thus, $I_{O,i}(t)$ is expressed in mA cm$^{-2}$, with a negative value during charging under the sign convention used here. The derivation of $S$, including the charge assigned to each subregion, is given in Supplementary Section S1.

In addition, the component shown in Fig. 2(b2) estimates the liquid-phase current density at time $t$ at the position $i = N + 1$ outside the observation points in the $z$-direction; similarly, it is represented by a three-layer NN as Eq. (3):

$$J_{O,bc}(t) = \sum_{k=1}^{K} b_{k,bc}\ \sigma\left(a_{k,1,bc} - a_{k,2,bc}\, t\right) + b_{K+1,bc}. \tag{3}$$

For training the NNs in Figs. 2(b1) and 2(b2), we optimize the connection weights by steepest descent with respect to the loss function defined in Eq. (4):

$$\begin{aligned}
\mathcal{L} &= \frac{1}{2M} \sum_{i,j,t \in \{x_{i,j}(t) \neq 1\}} \left(x_{i,j}(t) - y_{O,i}(t)\right)^2 \\
&+ \frac{1}{2M(1 - x_{th})} \sum_{i,j,t \in \{x_{i,j}(t) = 1\}} \int_{x_{th}}^{1} \left(x - y_{O,i}(t)\right)^2 dx \\
&+ \frac{A}{2} \sum_{t} \left(J_c - \sum_{i} I_{O,i}(t) - J_{O,\mathrm{bc}}(t)\right)^2 + \frac{B}{2} \sum_{i,t} \left(y_{O,i+1}(t) - y_{O,i}(t)\right)^2 \\
&+ \frac{C}{2} \sum_{i,t \in \{y_{O,i}(t) \leq 0.5\}} \left(0.5 - y_{O,i}(t)\right)^2 + \frac{C}{2} \sum_{i,t \in \{y_{O,i}(t) \geq 1\}} \left(1 - y_{O,i}(t)\right)^2.
\end{aligned} \tag{4}$$

The first term is a squared-error term for data fitting in the range $0.5 < x < x_{\mathrm{th}}$. The second term accounts for the fact that NLLS is not uniquely determined in the region $x \geq x_{\mathrm{th}}$, and is a squared-error term derived under the assumption that the data follow a uniform distribution. The summation over the lateral index $j$ corresponds to the operation of taking the ensemble average. The third term imposes current conservation by matching the sum of the LCO current densities over the modeled region and the liquid-phase current density at its interior boundary to the applied current density $J_c$ at the electrode edge. These current densities are all defined per unit cross-sectional area perpendicular to the $z$-direction; therefore, no additional area-conversion factor is required in this term. The fourth term enforces continuity of NLLS estimates between adjacent points in the $z$-direction. The fifth and sixth terms impose strong penalties when $y_{O,i}(t)$ falls below 0.5 or exceeds 1, respectively.

Using the method of steepest descent, we determine the connection weights and bias terms that minimize the loss function defined in Eq. (4), thereby training the network. The red lines shown in Figs. 2(b1) and 2(b2) represent the signal flow of error backpropagation during learning. During training, in addition to the error between the NLLS data $x_{i,k}(t)$ and the NN output $y_{O,i}(t)$, the gradients of the physics-informed regularization terms—such as current conservation and spatial continuity—are simultaneously propagated. Each connection weight and bias term is updated according to these gradient signals.

Figure 3(a) shows the estimated spatiotemporal NLLS distributions for samples with initial electrolyte concentrations of 0.3 M, 1 M, and 2 M. The charging current density was fixed to the experimental value $J_c = -6$ mA cm$^{-2}$. The sign of the charging current density is negative because it is flowing out from the electrode edge. The threshold for the weak-sensitivity region was set to $x_{\text{th}} = 0.85$. We also examined cases where the threshold was changed to $x_{\text{th}} = 0.9$, and found no essential differences in the estimated NLLS distributions, confirming that the proposed method is robust with respect to the threshold setting. The procedures used to determine the regularization parameters $A$, $B$, and $C$, as well as the number of hidden units $K$, and their specific values are provided in Supplementary Section S3.

## 2.6 Kirchhoff-MRF

In this study, we construct a Kirchhoff Markov random field (Kirchhoff-MRF) based on a simplified porous-electrode equivalent circuit. The relationships among the circuit variables prescribed by Kirchhoff's laws are encoded as probabilistic constraints, enabling the spatiotemporal distributions of latent electrochemical states to be inferred from the reconstructed NLLS maps.

Figure 4(a) shows the equivalent circuit used in this study. The electrode is discretized along the $z$-direction into spatial subregions of width $\Delta x$. Each subregion is treated as a control volume with an electrode cross-sectional area $A_{\perp}$ perpendicular to the $z$-direction. The lithium deintercalation/intercalation reaction in LCO at time $t$ and spatial index $i$ is represented by the charging or discharging of a capacitor.

The current density associated with the capacitor is denoted by $I_{\text{LCO},i}(t)$, and the equilibrium voltage across the capacitor, corresponding to the OCV of LCO, is denoted by $V_{\text{LCO},i}(t)$. The solid-liquid interfacial current density is denoted by $I_i(t)$, the liquid-phase current density along the $z$-direction by $J_i(t)$, the liquid-phase potential relative to the solid phase by $V_i(t)$, and the liquid-phase area-specific resistance across the spatial interval $\Delta x$ by $r_i(t)$.

The quantities $I_{\mathrm{LCO},i}(t)$, $I_i(t)$, and $J_i(t)$ are expressed in mA cm$^{-2}$. Here, $I_i(t)$ represents the volumetric interfacial reaction current integrated over the spatial interval $\Delta x$ and normalized by $A_\perp$. Consequently, $I_i(t)$ has the same units as $J_i(t)$, and these quantities can be directly related through the discrete current-conservation equation. The area-specific resistance $r_i(t)$ is expressed in kΩ cm$^2$, so that $r_i(t)J_i(t)$ represents the potential drop across the corresponding spatial interval in volts. As shown in Fig. 4(a), the solid-phase potential is taken as the reference potential. Accordingly, $V_i(t)$ and $V_{\mathrm{LCO},i}(t)$ are defined as the actual and equilibrium liquid-side potentials, respectively, relative to the solid phase, and both take negative values during charging. The signed interfacial overpotential entering the Butler-Volmer relation is therefore $V_i(t) - V_{\mathrm{LCO},i}(t)$. The positive directions of $I_{\mathrm{LCO},i}(t)$, $I_i(t)$, and $J_i(t)$ are defined by the corresponding arrows in Fig. 4(a). During charging, these current densities flow in the directions opposite to the arrows and consequently take negative values.

The LCO current density $I_{\mathrm{LCO},i}(t)$ and equilibrium potential $V_{\mathrm{LCO},i}(t)$ are obtained from the output of the physics-regularized three-layer neural network in the preceding stage of the analysis pipeline. Specifically, $I_{\mathrm{LCO},i}(t)$ is assigned the current density $I_{o,i}(t)$ calculated from the time derivative of the NN output $y_{o,i}(t)$ using Eq. (2). As described in Supplementary Section S1, the scaling coefficient in Eq. (2) converts the temporal variation in the NLLS of each spatial subregion into an area-normalized current density in mA cm$^{-2}$. The LCO equilibrium potential is calculated from the OCV curve in Supplementary Eq. (S8) as $V_{\mathrm{LCO},i}(t) = -E\left(y_{o,i}(t)\right)$, where the negative sign arises because the solid-phase potential is taken as the reference. The spatiotemporal distributions of $I_{\mathrm{LCO},i}(t)$ and $V_{\mathrm{LCO},i}(t)$ calculated from the estimated NLLS maps for initial electrolyte concentrations of 0.3 M, 1 M, and 2 M are shown in Figs. 3(b) and 3(c), respectively.

In the present reduced model, the potential drop in the solid phase is neglected because the electronic conductivity of the composite electrode is assumed to be sufficiently high compared with the effective ionic conductivity of the liquid phase. Ionic transport in the liquid phase is represented by an effective Ohmic relation. Electrolyte-species mass conservation, diffusion, migration, and concentration overpotential are not explicitly resolved.

The relationship between the interfacial current density $I_i(t)$ and the interfacial overpotential $V_i(t) - V_{\mathrm{LCO},i}(t)$ is described by a symmetric Butler–Volmer equation, as shown in Fig. 4(b):

$$I_i(t) = BV\left(V_i(t) - V_{\mathrm{LCO},i}(t)\right) = 2i_0 \sinh\left[\frac{F\{V_i(t) - V_{\mathrm{LCO},i}(t)\}}{2RT}\right]. \quad (5)$$

Here, $F$ is Faraday's constant, $R$ is the gas constant, and $T$ is the temperature. The parameter $i_0$ is the effective exchange current density in mA cm$^{-2}$. As described in Supplementary Section S1, $i_0$ is obtained by integrating the volumetric interfacial exchange current over the spatial interval $\Delta x$ and normalizing it by the electrode cross-sectional area $A_\perp$.

Within each initial electrolyte-concentration condition, $i_0$ was treated as a spatially and temporally constant effective parameter evaluated using the corresponding initial electrolyte concentration. The values of $i_0$ were set to $1.8040 \times 10^{-6}$, $3.2937 \times 10^{-6}$, and $4.6580 \times 10^{-6}$ mA cm$^{-2}$ for initial electrolyte concentrations of 0.3 M, 1 M, and 2 M, respectively. The derivation of these values is provided in Supplementary Section S1.

The effective liquid-phase conductivity $s_i(t)$ is related to the area-specific resistance $r_i(t)$ by

$$s_i(t) = \frac{\Delta x}{r_i(t)},$$

where $s_i(t)$, $\Delta x$, and $r_i(t)$ are expressed in mS cm$^{-1}$, cm, and k$\Omega$ cm$^2$, respectively. Therefore, $r_i(t) J_i(t)$ has units of volts when $J_i(t)$ is expressed in mA cm$^{-2}$.

In Fig. 4(a), $V_c$ and $J_c$ denote the experimentally measured charging voltage and charging current density, respectively, and are treated as known quantities. The liquid-phase current density $J_{N+1}(t)$, located outside the observation points on the electrode-interior side, is estimated from the auxiliary NN output $J_{o,\mathrm{bc}}(t)$ introduced in Section 2.5. The boundary conditions are therefore defined as

$$\begin{aligned} J_0(t) &= J_c, & J_{N+1}(t) &= J_{o,\mathrm{bc}}(t), \\ V_0(t) &= V_c, & V_{N+1}(t) &= V_N(t) - r_N(t) J_N(t), \\ r_0(t) &= r_1(t), & r_{N+1}(t) &= r_N(t). \end{aligned} \quad (6)$$

The boundary conditions for $r_i(t)$ correspond to free-end boundary conditions. The boundary condition for the liquid-phase potential is imposed by assuming that the potential drop across the final spatial interval is given by the same discrete Ohmic relation.

To estimate the latent state variables consistent with these governing relationships, we define the following Hamiltonian:

$$\mathcal{H} = H_\beta + H_\eta + H_\delta + H_\gamma + H_\phi + H_\psi, \qquad (7)$$

$$H_\beta = \beta \sum_{t=1}^{T} \sum_{i=1}^{N} \left( I_i(t) - I_{\mathrm{LCO},i}(t) \right)^2,$$

$$H_\eta = \eta \sum_{t=1}^{T} \sum_{i=1}^{N} \left( V_i(t) - V_{\mathrm{LCO},i}(t) - BV^{-1}\left(I_i(t)\right) \right)^2,$$

$$H_\delta = \delta \sum_{t=1}^{T} \sum_{i=1}^{N} \left( J_i(t) - J_{i-1}(t) + I_i(t) \right)^2,$$

$$H_\gamma = \gamma \sum_{t=1}^{T} \sum_{i=1}^{N} \left( V_{i+1}(t) - V_i(t) + r_i(t) J_i(t) \right)^2,$$

$$H_\phi = \phi \sum_{t=1}^{T} \sum_{i=1}^{N} \left( r_{i+1}(t) - r_i(t) \right)^2,$$

$$H_\psi = \psi \sum_{t=1}^{T} \sum_{i=1}^{N} \left( r_i(t) - \bar{r} \right)^2.$$

The first term represents the residual between the LCO current density estimated by the three-layer NN and the interfacial current density. The second term represents the nonlinear current-voltage relationship at the solid-liquid interface through the inverse Butler-Volmer relation. Because this term contains $V_{\mathrm{LCO},i}(t)$ obtained from the estimated NLLS, it also provides a potential-related likelihood term. The third term represents the discrete current-conservation relation and corresponds to Kirchhoff's current law. Because $I_i(t)$ is the reaction current density integrated over the spatial interval $\Delta x$, an additional factor of $\Delta x$ is not required in this term. If a volumetric reaction current density $I_i^{\mathrm{vol}}(t)$, expressed in mA cm$^{-3}$, were used instead, the corresponding term would contain $\Delta x I_i^{\mathrm{vol}}(t)$. The fourth term represents the liquid-phase Ohmic voltage drop and corresponds to Kirchhoff's voltage law. The fifth term imposes spatial continuity on the liquid-phase area-specific resistance, and the sixth term constrains $r_i(t)$ around the reference area-specific resistance $\bar{r}$, which is determined from the initial electrolyte concentration.

The Hamiltonian is constructed as a weighted linear combination of these terms using the hyperparameters $\beta$, $\eta$, $\delta$, $\gamma$, $\phi$, and $\psi$. Because the Hamiltonian includes the nonlinear Butler–Volmer relationship, it is not a quadratic function. The inverse Butler–Volmer function is used in the second term to facilitate the locally Gaussian approximation employed in the approximate Gibbs sampler described in Supplementary Section S2.

Based on this Hamiltonian, the Kirchhoff-MRF is expressed as the following high-dimensional probability distribution:

$$P(J, I, V, r, I_{\mathrm{LCO}}, V_{\mathrm{LCO}} \mid \beta, \eta, \delta, \gamma, \phi, \psi, \bar{r})$$

$$= P(I_{\mathrm{LCO}} \mid I, \beta) P(V_{\mathrm{LCO}} \mid V, I, \eta) P(I \mid J, \delta) P(J, V \mid r, \gamma) P(r \mid \phi, \psi, \bar{r}). \qquad (8)$$

The individual probability-density terms are

$$P(I_{\mathrm{LCO}} \mid I, \beta) \propto \prod_{t=1}^{T} \exp\left(-\beta \sum_{i=1}^{N} \left(I_i(t) - I_{\mathrm{LCO},i}(t)\right)^2\right), \qquad (9)$$

$$P(V_{\mathrm{LCO}} \mid V, I, \eta) \propto \prod_{t=1}^{T} \exp\left(-\eta \sum_{i=1}^{N} \left(V_i(t) - V_{\mathrm{LCO},i}(t) - BV^{-1}\left(I_i(t)\right)\right)^2\right), \qquad (10)$$

$$P(I \mid J, \delta) \propto \prod_{t=1}^{T} \exp\left(-\delta \sum_{i=1}^{N} \left(J_i(t) - J_{i-1}(t) + I_i(t)\right)^2\right), \qquad (11)$$

$$P(J, V \mid r, \gamma) \propto \prod_{t=1}^{T} \exp\left(-\gamma \sum_{i=1}^{N} \left(V_{i+1}(t) - V_i(t) + r_i(t) J_i(t)\right)^2\right), \qquad (12)$$

$$P(r \mid \phi, \psi, \bar{r}) \propto \prod_{t=1}^{T} \exp\left(-\phi \sum_{i=1}^{N} \left(r_{i+1}(t) - r_i(t)\right)^2 - \psi \sum_{i=1}^{N} \left(r_i(t) - \bar{r}\right)^2\right). \qquad (13)$$

The dependency structure among the variables is represented by the hierarchical Bayesian model shown in Fig. 4(c). Under the quasi-static approximation, the latent states at different measurement times are treated as conditionally independent. Subject to the boundary conditions in Eq. (6), approximate marginal posterior distributions of $I_i(t)$, $J_i(t)$, $V_i(t)$, and $r_i(t)$ are obtained using the approximate Gibbs sampling procedure described in Supplementary Section S2.

# 3. Results

## 3.1 Estimation of internal battery states from operando microscopic spectroscopy data

In this section, we apply the proposed analysis pipeline shown in Fig. 1(c) to operando $\mu$-XAFS hyperspectral datasets acquired for three LIB composite cathode samples with different initial electrolyte concentrations ($LiPF_6$: 0.3 M, 1 M, and 2 M), and elucidate the spatiotemporal dynamics of internal battery states during charging.

Under the conditions described in Section 2.1, the charging current density was fixed at $J_c = -6\ \mathrm{mA\,cm^{-2}}$, and data up to the point where the charging voltage reached $V_c = -4.3\ \mathrm{V}$ were analyzed. Note that in this model the solid phase is taken as ground and the liquid-phase potential is defined relative to it; the positive direction of current is also defined as the direction indicated by the arrows in Fig. 4(a). Accordingly, $V_c$ and $J_c$ take negative values under the present sign convention.

By applying the methods described in Sections 2.3 and 2.5 to the Co K-edge XAFS spectra obtained by operando microscopic measurements, we estimated the spatiotemporal NLLS distribution including the weak-sensitivity region (Fig. 3(a)), and calculated the lithium deintercalation/intercalation current $I_{LCO}$ and the OCV $V_{LCO}$ in LCO (Figs. 3(b) and 3(c)). Furthermore, by applying the Kirchhoff-MRF, we estimated the spatiotemporal distributions of the solid-liquid interfacial current density $I$, liquid-phase current density $J$, liquid-phase potential $V$, and effective liquid-phase conductivity $s = \Delta x/r$, where $\Delta x$ is the spatial sampling interval and $r$ is the liquid-phase ohmic area-specific resistance. For visualization in Fig. 5, the interfacial current density $I$, which is expressed in mA cm$^{-2}$, was converted to the corresponding volumetric reaction current density $I^{vol} = I/\Delta x$, expressed in mA cm$^{-3}$.

Figure 5 shows the temporal evolution of the spatial distributions of the estimated internal states. The first column presents the volumetric reaction current density $I^{vol} = I/\Delta x$, whereas the remaining columns present the liquid-phase current density $J$, liquid-phase potential $V$, and effective liquid-phase conductivity $s$. To facilitate comparison of the spatial profiles at different times, the profiles are vertically offset, and the elapsed charging time corresponding to each profile is indicated on the right. The horizontal guide lines and numerical annotations indicate the reference levels used for the vertically offset profiles and selected boundary values, as specified in the figure caption.

### 3.1.1 Case of initial electrolyte concentration 0.3 M

Figure 5(a) shows the estimation results for the sample with an initial electrolyte concentration of 0.3 M. Focusing on the distribution of the volumetric interfacial reaction current density $I^{vol}$, a negative dip is observed to move from the electrode edge (distance 0) toward the interior as time progresses, as indicated by the gray arrows in the figure. This negative volumetric reaction-current-density dip corresponds to Li deintercalation from LCO, and its motion reflects the progression of the charging reaction from the electrode edge toward the electrode interior.

In the distribution of the liquid-phase conductivity $s$, an increase in conductivity is observed near the electrode edge immediately after the start of charging, and this high-conductivity region expands inward with time. Simultaneously with this conductivity increase, the spatial gradient of the liquid-phase potential $V$ gradually becomes gentler, suggesting a reduction in ohmic loss associated with ionic transport in the liquid phase.

These results indicate that, as the ohmic loss decreases, the electrode-edge potential that drives the reaction is sufficiently transmitted into the electrode interior, thereby allowing the charging reaction to propagate inward. Moreover, the gradual propagation of the reaction from the electrode edge, which serves as the ion outlet, suggests that ionic transport may play a dominant role in governing the overall reaction kinetics in this electrode. This behavior implies that electronic conduction through the conductive additive or active material, as well as interfacial charge-transfer reactions, are unlikely to be the primary rate-limiting steps under the present conditions.

### 3.1.2 Cases of initial electrolyte concentration 1 M and 2 M

Figures 5(b) and 5(c) show the estimation results for samples with initial electrolyte concentrations of 1 M and 2 M, respectively. Under these conditions, behavior contrasting with the 0.3 M case was observed.

In the distribution of the volumetric interfacial reaction current density $I^{vol}$, the negative current-density region remains localized near the electrode edge throughout charging, and no clear inward migration with time is observed, as indicated by the gray arrows. That is, the charging reaction is localized near the electrode edge and its progression into the interior is suppressed.

Moreover, the liquid-phase conductivity $s$ decreases monotonically with time after the start of charging, and no local conductivity increase such as that observed in the 0.3 M case is detected. Correspondingly, the spatial gradient of the liquid-phase potential $V$ increases, indicating an increase in ohmic loss associated with ionic transport in the liquid phase.

Taken together, these results suggest that, for initial electrolyte concentrations of 1 M and 2 M, increased ohmic loss in the liquid phase hinders potential transmission into the electrode interior, thereby confining the charging reaction to the vicinity of the electrode edge.

## 3.2 Comparison between estimated electrolyte concentration distributions and operando measurements

In this section, we estimate the effective electrolyte concentration distribution from the reconstructed liquid-phase conductivity and assess its qualitative consistency with independent operando measurements.

### 3.2.1 Concentration estimation based on the conductivity–electrolyte concentration relationship

A previous study [15,40,41] reported an empirical relationship between the liquid-phase conductivity and electrolyte concentration in LIB cathodes employing a $LiPF_6$ electrolyte (Supplementary Eq. (S9)). Figure 6(a) shows this relationship as a function of salt concentration. The conductivity exhibits a non-monotonic behavior: it takes a maximum around $\sim 1\ \mathrm{M}$ and decreases toward both lower and higher concentrations.

In this study, we estimated the electrolyte concentration distribution in the electrode at each time by substituting the estimated conductivity $s$ (Fig. 5) into this relationship. Because this relationship yields two concentration solutions for a single conductivity value, we selected the solution that preserves temporal continuity from the initial electrolyte concentration.

Figures 6(b)–6(d) show the temporal evolution of the estimated electrolyte concentration distributions for initial $LiPF_6$ concentrations of 0.3 M, 1 M, and 2 M, respectively, together with the corresponding changes in charging voltage and charging capacity.

### 3.2.2 Characteristics of the estimated concentration distributions

For the initial concentration of 0.3 M in Fig. 6(b), a sharp concentration profile forms near the electrode edge immediately after the start of charging, and the profile broadens toward the interior with time. This progression of the concentration distribution is consistent with the inward propagation of the charging reaction shown in Section 3.1.1. It took approximately 105 min for the charging voltage to reach the upper limit, during which the charging capacity reached approximately $25\ \mathrm{mAh/g}$. No saturation of the concentration or charging voltage was observed in the late stage of charging.

In contrast, for initial concentrations of 1 M and 2 M in Figs. 6(c) and 6(d), a concentration distribution spreading across the entire electrode forms immediately after charging starts, followed by an overall increase in concentration. The charging voltage reached the upper limit after approximately 64 min for 1 M and approximately 28 min for 2 M, and the charging capacities were lower than that for 0.3 M in both cases. No saturation of the concentration or charging voltage was observed in the late stage of charging.

### 3.2.3 Comparison with operando X-ray transmission measurements

To provide an independent qualitative consistency check, we compared the inferred electrolyte profiles with operando X-ray transmission measurements obtained using a model electrode with a geometry similar to that employed in the $LiPF_6$ experiments. $LiAsF_6$ was used instead of $LiPF_6$ because the high X-ray attenuation of arsenic enables spatial changes in electrolyte concentration to be visualized as changes in X-ray transmission contrast. In this experiment, the initial electrolyte concentration was 1 M, and the cell was charged for 45 min at a current density of 6 mA $cm^{-2}$. Further experimental details are provided in Section 2.2. Figure 7 shows the temporal evolution of the electrolyte concentration distribution inferred from the differences between transmission images acquired before and during charging, together with the corresponding charging voltage and capacity.

For the $LiAsF_6$ electrolyte, a spatially extended concentration change appeared across the electrode shortly after charging began, followed by a further increase throughout the electrode. This behavior qualitatively resembles the temporal evolution of the effective electrolyte concentration profiles inferred for the 1 M and 2 M $LiPF_6$ conditions in Figs. 6(c) and 6(d).

In the $LiAsF_6$ experiment, however, the concentration increase tended to saturate during the late stage of charging (35–45 min). The charging voltage shown in the right panel of Fig. 7 also exhibited saturation over approximately the same period. Comparable late-stage saturation was not observed in the inferred profiles for the 1 M and 2 M $LiPF_6$ conditions shown in Figs. 6(c) and 6(d).

Except for the late-stage saturation, the 1 M $LiAsF_6$ experiment and the inferred 1 M and 2 M $LiPF_6$ profiles exhibited a similar spatially extended concentration change across the electrode.

# 4. Discussion

The liquid-phase conductivity reconstructed from the operando $\mu$-XAFS measurements was converted into an effective electrolyte concentration using the experimentally established conductivity–concentration relationship shown in Fig. 6(a). For initial electrolyte concentrations of 1 M and 2 M, the reconstructed concentration profiles qualitatively agree with the independent operando X-ray transmission measurements using a $LiAsF_6$ electrolyte (Figs. 6(c), 6(d), and 7). The spatial resolution of the operando X-ray transmission imaging was 1.3 μm, which was substantially smaller than the characteristic spatial extent of the observed concentration variations. Thus, the spatially extended profiles observed in the transmission images are unlikely to be attributable solely to instrumental blurring. However, this comparison should be interpreted as a qualitative consistency check rather than quantitative validation because the electrolyte chemistry differs and the independent measurement was performed only at an initial concentration of 1 M. $LiPF_6$ and $LiAsF_6$ differ in anion species and may therefore exhibit different transport properties, including diffusion coefficients and solvation structures. The late-stage saturation of electrolyte concentration and charging voltage observed in the $LiAsF_6$ transmission experiment was not reproduced in the $LiPF_6$ results. This difference may arise from differences in electrolyte chemistry, measurement duration, or electrochemical conditions; however, its origin cannot be determined from the present comparison. For the 0.3 M condition, the framework reconstructs an effective electrolyte concentration profile in which the region of elevated concentration extends inward from the electrode edge (Fig. 6(b)). This spatial evolution is consistent with the experimentally observed inward propagation of the charging reaction and is therefore electrochemically plausible. Independent experimental evaluation of the characteristic concentration profile inferred for the 0.3 M condition remains an important subject for future work.

In the previous study by Nakamura et al. [15], the concentration-dependent reaction behavior was qualitatively attributed to changes in electrolyte concentration within the electrode; however, its spatiotemporal evolution was not directly resolved. The present framework provides a model-based, spatially resolved means of evaluating this hypothesis. Specifically, the effective electrolyte conductivity is reconstructed from the spectroscopic data under the assumptions of the Kirchhoff–MRF model, and the corresponding effective concentration is subsequently inferred using the empirical conductivity–concentration relationship. At an initial concentration of 0.3 M, the charging reaction propagates from the electrode edge toward the interior, whereas at 1 M and 2 M, it remains localized near the electrode edge. For the 0.3 M condition, the inferred effective electrolyte concentration initially increases near the electrode edge, after which the region of elevated concentration extends toward the electrode interior.

In contrast, the 1 M and 2 M conditions exhibit spatially extended concentration changes across the electrode from an earlier stage of charging. This concentration-dependent contrast in the reconstructed profiles is consistent with a mechanism involving the non-monotonic dependence of ionic conductivity on electrolyte concentration, as shown in Fig. 6(a). Although electrolyte-species mass conservation and diffusion are not explicitly modeled, the reconstructed profile provides a spatially resolved representation consistent with inward electrolyte redistribution.

The Kirchhoff-MRF employed here is based on a simplified porous-electrode equivalent circuit designed to balance the identifiability of latent states with computational tractability. Solid-phase resistance is neglected because the present model electrode is thin and contains a sufficient amount of conductive additive; however, this approximation may require reconsideration for thicker electrodes or electrodes with lower electronic conductivity. Ionic transport is represented by an effective Ohmic relation, without explicitly resolving electrolyte-species mass conservation, diffusion, migration, or concentration overpotential. This reduced description is considered suitable for characterizing the principal spatial trends under the moderate-rate conditions examined here, whereas an extended transport model may be required at higher rates or when steep concentration gradients develop. Furthermore, the interfacial reaction is described using a symmetric Butler-Volmer relation. Within each initial-concentration condition, the exchange current density is treated as a spatially and temporally constant effective parameter; therefore, its local dependence on the Li concentration in LCO and the electrolyte concentration, as well as interfacial heterogeneity, is not explicitly considered. Despite these simplifications, the reconstructed fields reveal systematic concentration-dependent spatiotemporal trends that are consistent with the experimentally observed reaction behavior.

The proposed Kirchhoff-MRF includes multiple hyperparameters that control the relative contributions of the likelihood and regularization terms and can affect both numerical stability and the physical plausibility of the reconstructed fields. In the present study, these hyperparameters were empirically selected to satisfy predefined numerical and physical admissibility criteria, as detailed in Supplementary Section S3. A more systematic, data-driven approach, such as marginal-likelihood maximization or a hierarchical Bayesian formulation with hyperpriors, could further improve the reproducibility of hyperparameter selection. The local sensitivity analysis presented in Supplementary Section S4 further shows that the principal concentration-dependent trends are preserved for parameter values adjacent to the adopted hyperparameter pair. Although the present analysis focuses on a one-dimensional reaction-propagation geometry, the general formulation could, in principle, be extended to higher spatial dimensions by reformulating the equivalent-circuit topology and the associated spatial interactions. Moreover, NLLS estimation using the physics-regularized neural network and ion-transport inference using the Kirchhoff-MRF are performed sequentially in the present framework. A fully joint formulation within a unified probabilistic objective, potentially implemented using variational-inference approaches such as that described in Ref. [43], could enable simultaneous inference of the latent NLLS maps and transport variables and may improve their identifiability and internal consistency. From a machine-learning perspective, the proposed pipeline constitutes a structured probabilistic framework that incorporates physical constraints into latent-state inference. This framework may provide a basis for extension to other active materials, battery chemistries, and electrochemical systems.

## Data availability

The operando $\mu$-XAFS datasets and X-ray transmission images are available from the corresponding author upon reasonable request.

## Code availability

Custom code implementing the physics-regularized neural network and Kirchhoff-MRF inference pipeline is available from the corresponding author upon reasonable request.

## Acknowledgements

This work was supported by JST CREST (JPMJCR1861, JPMJCR2431 and JPMJCR2432).

## Author contributions

N.W. and T.A. developed the analysis pipeline and conducted the data analysis. Y.K. and K.A. acquired the experimental data. M.M. and I.A. contributed to project administration. T.A. conceived and supervised the research. All authors contributed to the interpretation of results and the preparation of the manuscript.

## Competing interests

The authors declare no competing interests.

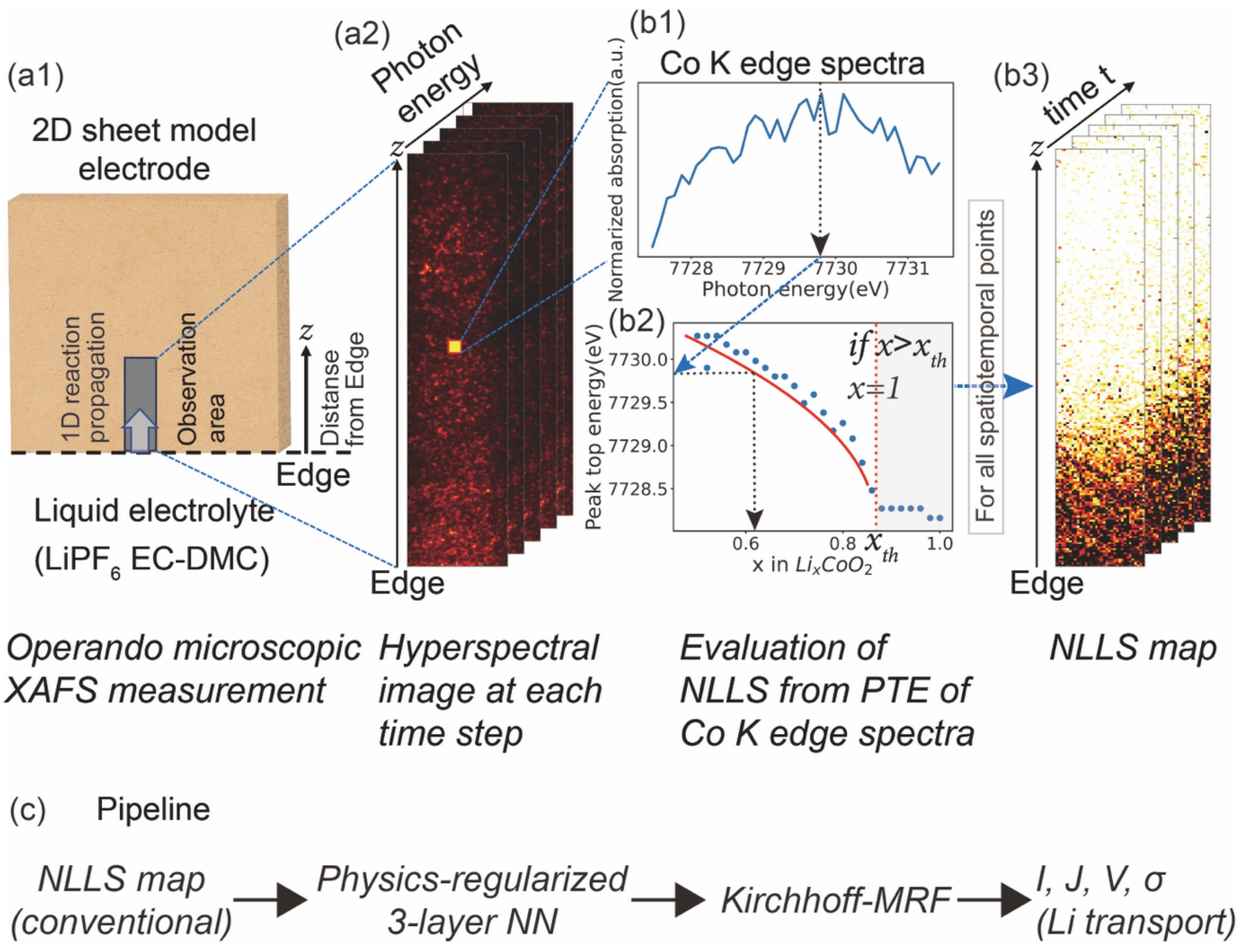

**Figure 1. Hyperspectral data obtained by Co K-edge $\mu$-XAFS measurements and the proposed analysis pipeline.**

**(a)** Schematic of operando micro-XAFS measurements for a two-dimensional sheet-type model electrode. **(a1)** Schematic illustration of the model electrode. The electrode edge is maintained at an equipotential, and the charge-discharge reaction proceeds one-dimensionally along the $z$-axis indicated in the figure. **(a2)** Example of a hyperspectral image acquired at a given time. Hyperspectral images are acquired at each time point during the progression of the charge-discharge reaction. **(b)** Procedure for constructing an NLLS map using the conventional method. **(b1)** Example of a Co K-edge XAFS spectrum at a given spatiotemporal point and the corresponding PTE. **(b2)** Relationship between PTE and NLLS obtained from standard samples, together with a quadratic regression curve. For the PTE obtained at each spatiotemporal point, NLLS is assigned based on the regression curve when $x < x_{\mathrm{th}}$. In contrast, when $x > x_{\mathrm{th}}$ (weak-sensitivity region), NLLS cannot be uniquely determined; therefore, the representative value of 1 is assigned. **(b3)** Example of NLLS maps at each time point obtained by the above procedure. **(c)** Proposed analysis pipeline. For the NLLS map obtained by the conventional method, a three-layer NN is regressed under physics-informed regularization to estimate NLLS in the weak-sensitivity region (see Fig. 2 for details). In addition, by using a Kirchhoff Markov random field (Kirchhoff-MRF) based on a simplified porous-electrode model, internal states such as interfacial current density, electrolyte current, potential, and conductivity in the Li-ion transport process are estimated (see Fig. 4 for details).

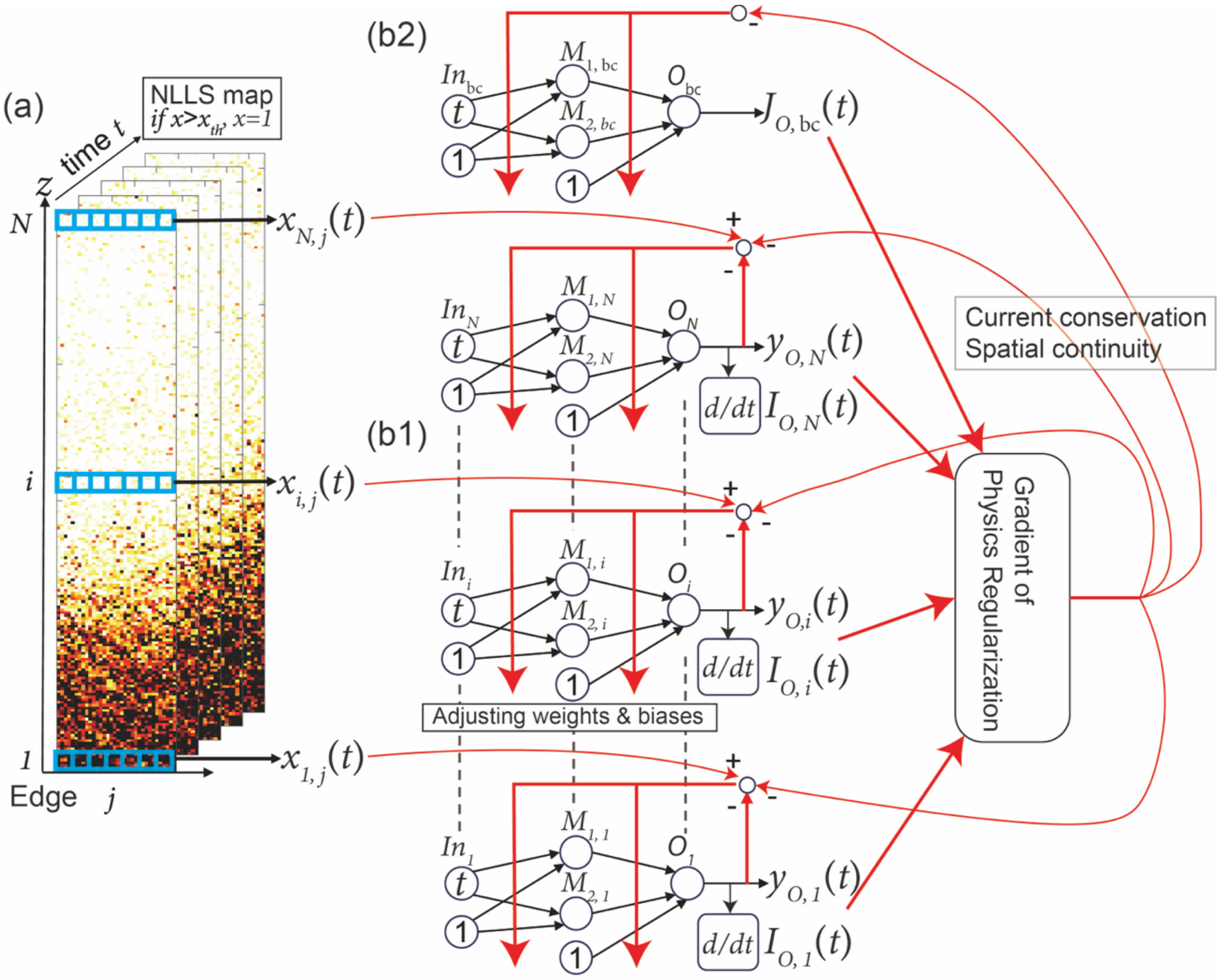

**Figure 2. Data structure of the spatiotemporal NLLS map and physics-regularized three-layer neural network.**

**(a)** Data structure of the spatiotemporal NLLS map. The NLLS at time $t$, pixel index $i$ along the $z$-axis, and lateral pixel index $j$ is defined as $x_{i,j}(t)$. **(b)** Structure of the physics-regularized three-layer NN. **(b1)** For each pixel index $i$ along the $z$ direction ($i = 1, \dots, N$), the network consists of an input-layer cell $In_i$ that outputs time $t$, $K$ hidden-layer cells $M_{1,i}$ to $M_{K,i}$ (in the figure, $K = 2$) with a sigmoid activation function, and an output-layer cell $O_i$ with a linear activation function. Note that ① is a cell that outputs the constant 1, corresponding to the bias term. The output of $O_i$ is $y_{O,i}(t)$, and its time derivative is interpreted as the current $I_{O,i}(t)$ corresponding to Li deintercalation/intercalation from/into LCO. **(b2)** Network representing the liquid-phase current density at time $t$ at $i = N + 1$, which is located outside the observation points in the $z$-direction, expressed using the same three-layer NN structure. Black lines indicate the forward signal flow of the network, and red lines indicate the signal flow during training. The connection weights are updated by considering not only the error between the NLLS data $x_{i,k}(t)$ and the output $y_{O,i}(t)$, but also the gradients of the regularization terms based on NLLS spatial continuity and current conservation.

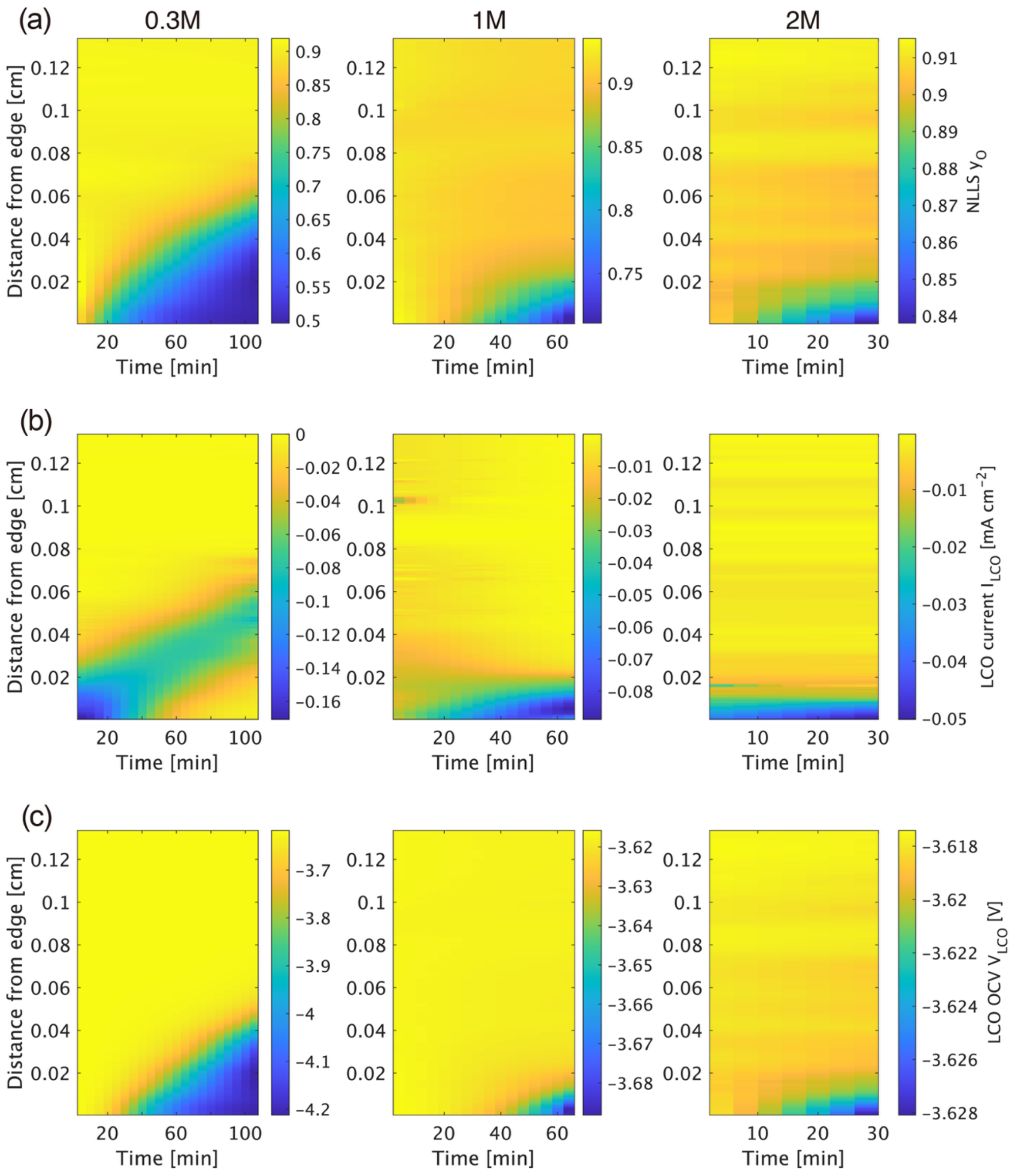

**Figure 3. Estimated spatiotemporal NLLS map including the weak-sensitivity region, and estimated LCO current density and open-circuit voltage.**

**(a)** Estimated spatiotemporal NLLS maps including the weak-sensitivity region, plotted from the trained NN output $y_{O,i}(t)$ for initial electrolyte concentrations of 0.3 M, 1 M, and 2 M. **(b)** Spatiotemporal map of the LCO current density. The current density $I_{O,i}(t)$, obtained as the time derivative of the NN output $y_{O,i}(t)$ shown in (a), is assigned as the evaluated LCO current density $I_{LCO,i}(t)$. **(c)** Spatiotemporal map of the LCO OCV. By applying the OCV curve in Supplementary Eq. (S8) as $V_{\mathrm{LCO},i}(t) = -E\left(y_{o,i}(t)\right)$, the LCO OCV $V_{LCO,i}(t)$ is calculated and plotted.

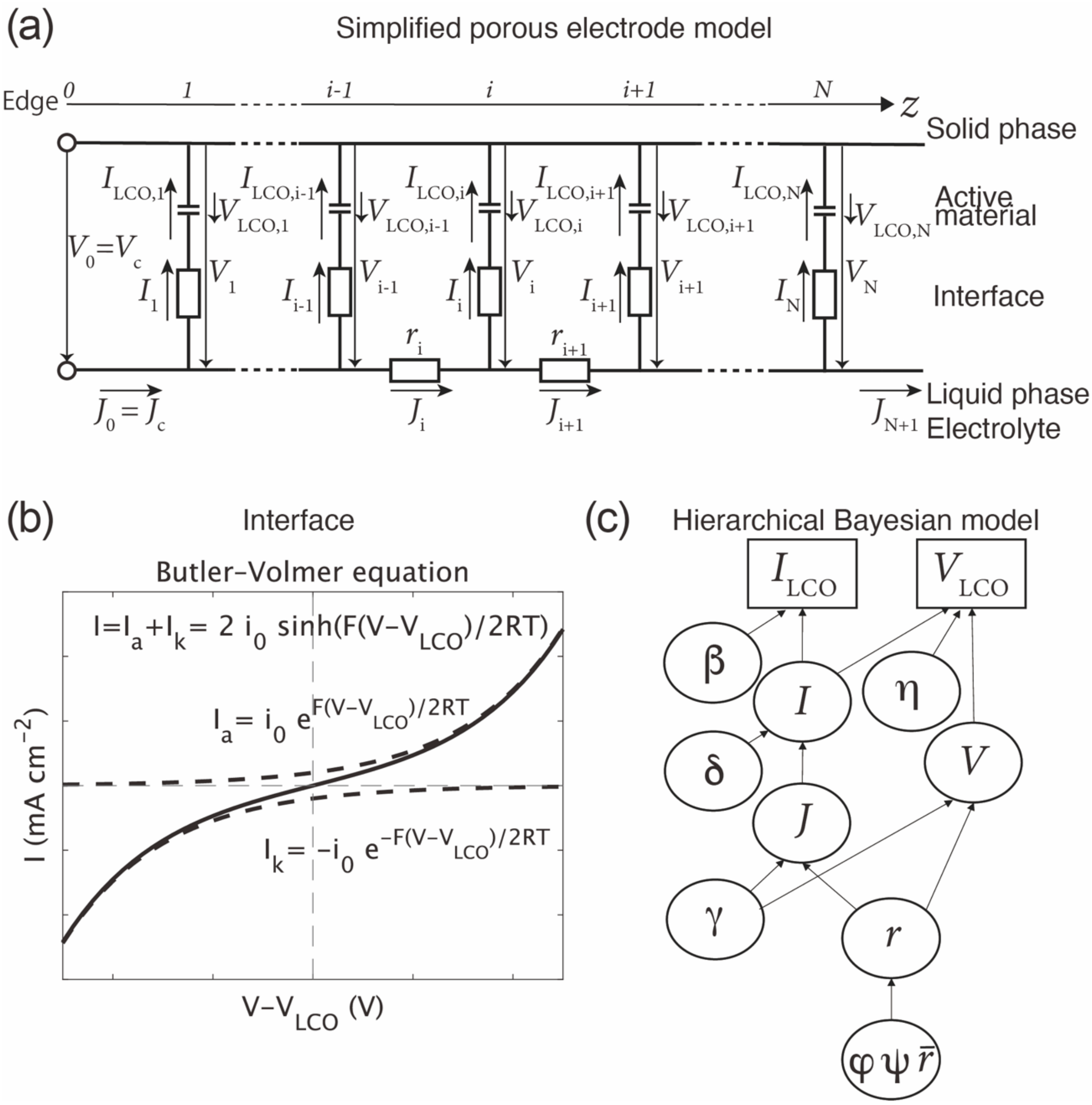

**Figure 4. Kirchhoff Markov random field (Kirchhoff-MRF) for inferring Li-ion transport processes.**

**(a)** Equivalent circuit based on the simplified porous-electrode model. The electrode is discretized along the $z$-direction into spatial subregions of width $\Delta x$. Shown are the LCO current density $I_{\mathrm{LCO},i}$, LCO equilibrium potential $V_{\mathrm{LCO},i}$, solid-liquid interfacial current density $I_i$, liquid-phase current density $J_i$, liquid-phase potential $V_i$, liquid–phase area–specific resistance $r_i$, charging current density $J_c$, charging voltage $V_c$, and liquid–phase current density $J_{N+1}$ outside the observation region. The solid–phase potential is taken as the reference potential. Accordingly, $V_i$ and $V_{\mathrm{LCO},i}$ are defined as liquid–side potentials relative to the solid phase. The arrows indicate the positive directions of the current densities and the reference polarities of the potentials. The quantities $I_{\mathrm{LCO},i}$, $I_i$, and $J_i$ are expressed in mA cm$^{-2}$, where $I_i$ represents the volumetric interfacial reaction integrated over $\Delta x$ and normalized by the electrode cross–sectional area. The area–specific resistance $r_i$ is expressed in kΩ cm$^2$. **(b)** Symmetric Butler–Volmer relationship between the interfacial current density $I_i$ and the signed interfacial overpotential $V_i - V_{\mathrm{LCO},i}$. The parameter $i_0$ denotes the effective exchange current density. The vertical axis is expressed in mA cm$^{-2}$. **(c)** Hierarchical Bayesian representation of the Kirchhoff–MRF. The dependency relationships among the observed quantities, latent state variables, and hyperparameters are constructed from the current–conservation relation, the liquid–phase Ohmic relation, the Butler–Volmer relation, and spatial regularization of the area–specific resistance.

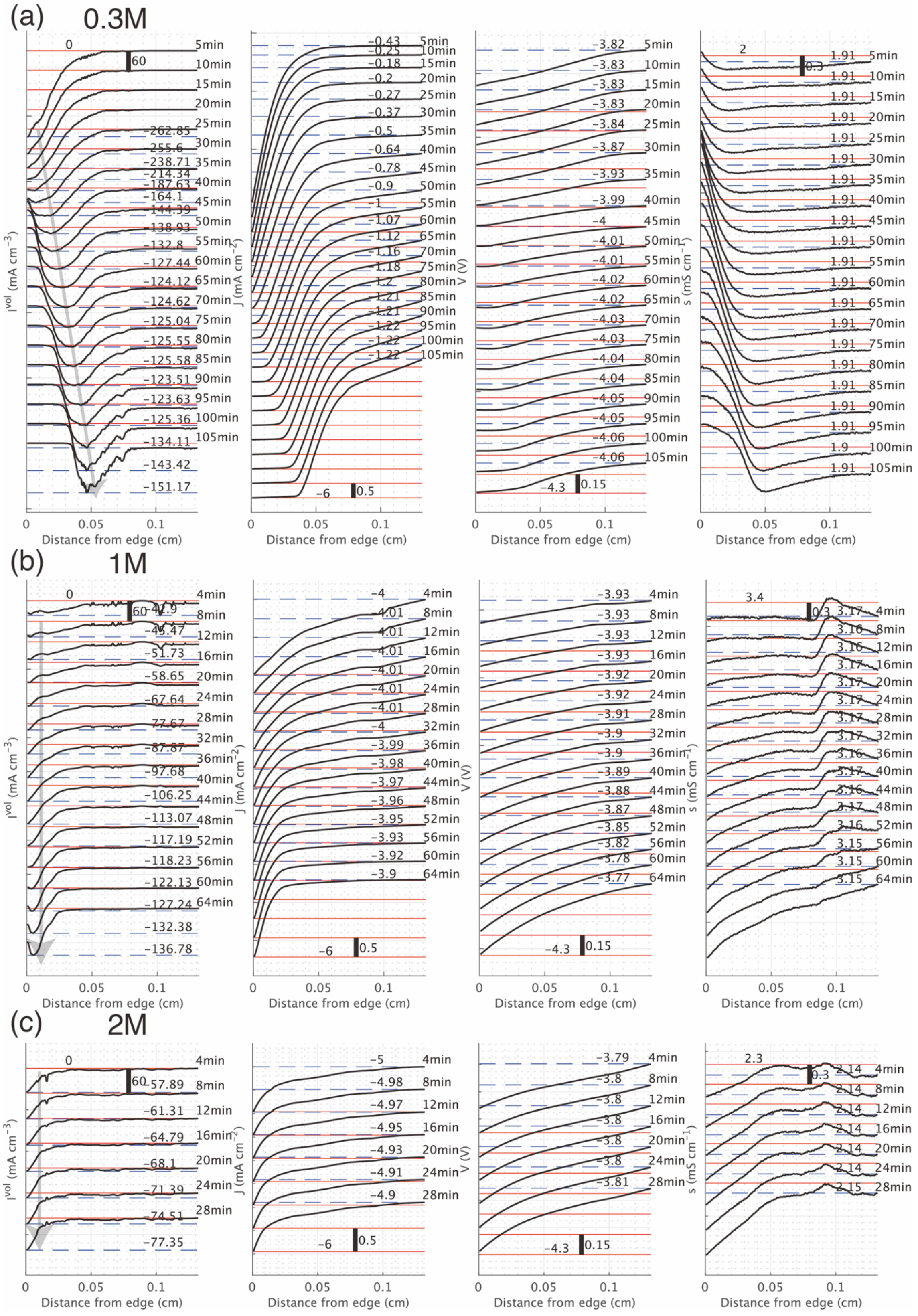
(a)
0.3M
(b)
1M
(c)
2M
$I^{vol}$ (mA cm$^{-3}$)
J (mA cm$^{-2}$)
V (V)
s (mS cm$^{-1}$)
Distance from edge (cm)

**Figure 5. Analysis results of operando micro-spectroscopy data capturing the charging process using the proposed pipeline.**

Time evolution of the spatial distributions of the volumetric interfacial reaction current density $I^{vol} = I/\Delta x$, liquid-phase current density $J$, liquid-phase potential $V$, and effective liquid-phase conductivity $s$, as estimated by the proposed analysis pipeline. Here, $\Delta x$ is the spatial sampling interval, and $I$ is the interfacial reaction current density integrated over $\Delta x$ and normalized by the electrode cross-sectional area. Accordingly, $I$ and $I^{vol}$ are expressed in mA cm$^{-2}$ and mA cm$^{-3}$, respectively. Results are shown for initial electrolyte concentrations of **(a)** 0.3 M, **(b)** 1 M, and **(c)** 2 M. The spatial profiles at successive times are vertically offset for clarity, and the corresponding elapsed charging times are indicated on the right. Red horizontal lines indicate the reference levels used for the vertically offset profiles. Light-blue dashed lines indicate the values at the electrode-interior boundary. The annotated numerical values correspond to the respective reference or boundary values. Red horizontal lines indicate the reference levels used for displaying the vertically offset profiles. In the $I^{vol}$ panels, the light-blue dashed lines indicate the minimum values of the negative dips in the spatial profiles. In the $J$, $V$, and $s$ panels, the light-blue dashed lines indicate the values at the electrode-interior boundary. The annotated numerical values correspond to the respective red-line reference levels or light-blue dashed-line values.

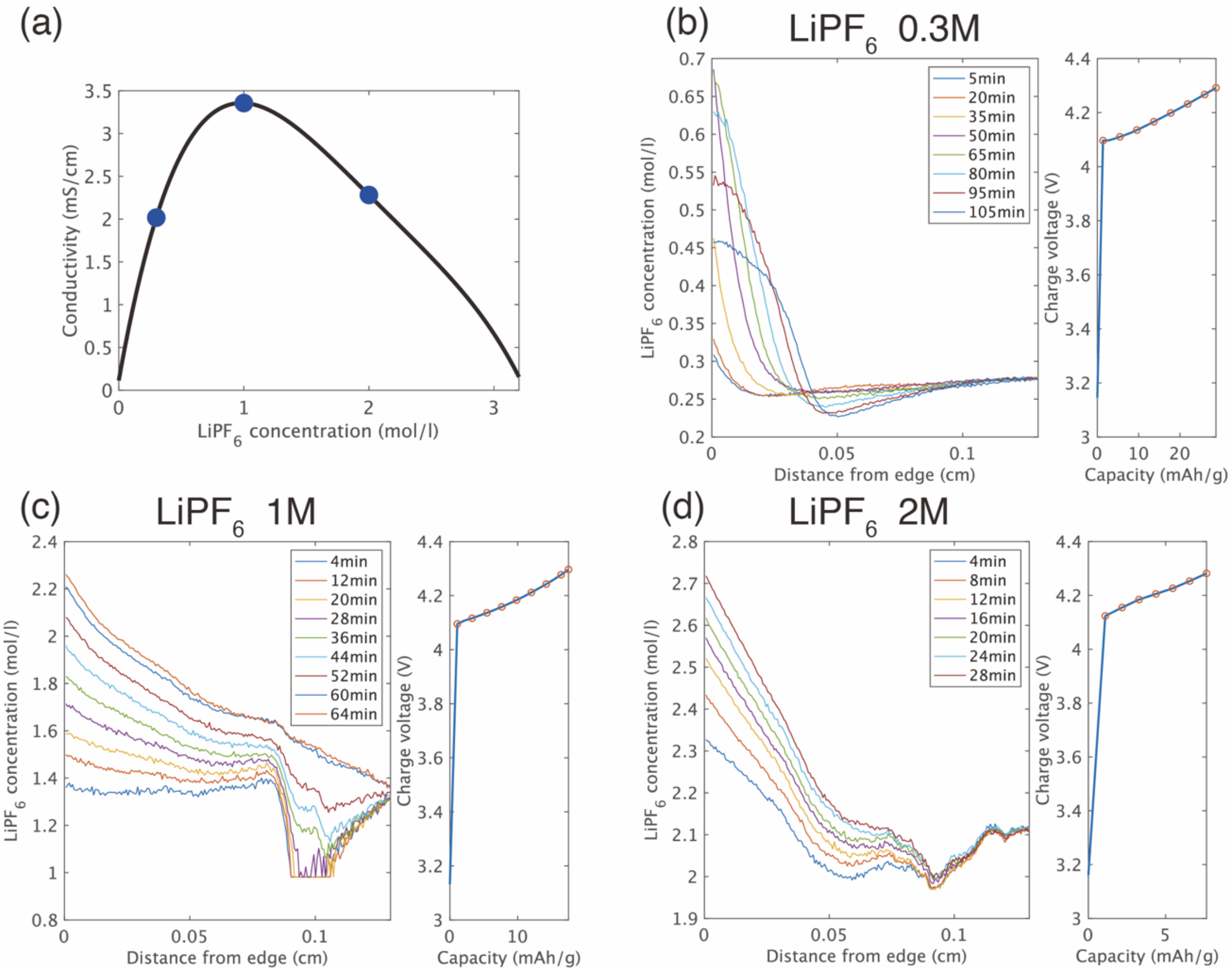

**Figure 6. Results of predicting the electrolyte concentration from the estimated liquid-phase ohmic conductivity.**

**(a)** Relationship between ionic conductivity and electrolyte concentration in an electrode device using $LiPF_6$ electrolyte, as expressed by Supplementary Eq. (S9). Blue circles indicate the conductivities under initial electrolyte concentrations of 0.3 M, 1 M, and 2 M. The left panels in **(b)–(d)** show the time evolution of the spatial distribution of electrolyte concentration evaluated for sample electrode devices employing $LiPF_6$, by applying the relationship in (a) to the estimated conductivity $s$ in Fig. 5. The initial electrolyte concentrations are **(b)** 0.3 M, **(c)** 1 M, and **(d)** 2 M. The right panels in **(b)–(d)** show the time evolution of the charging voltage and charge capacity for each electrode device.

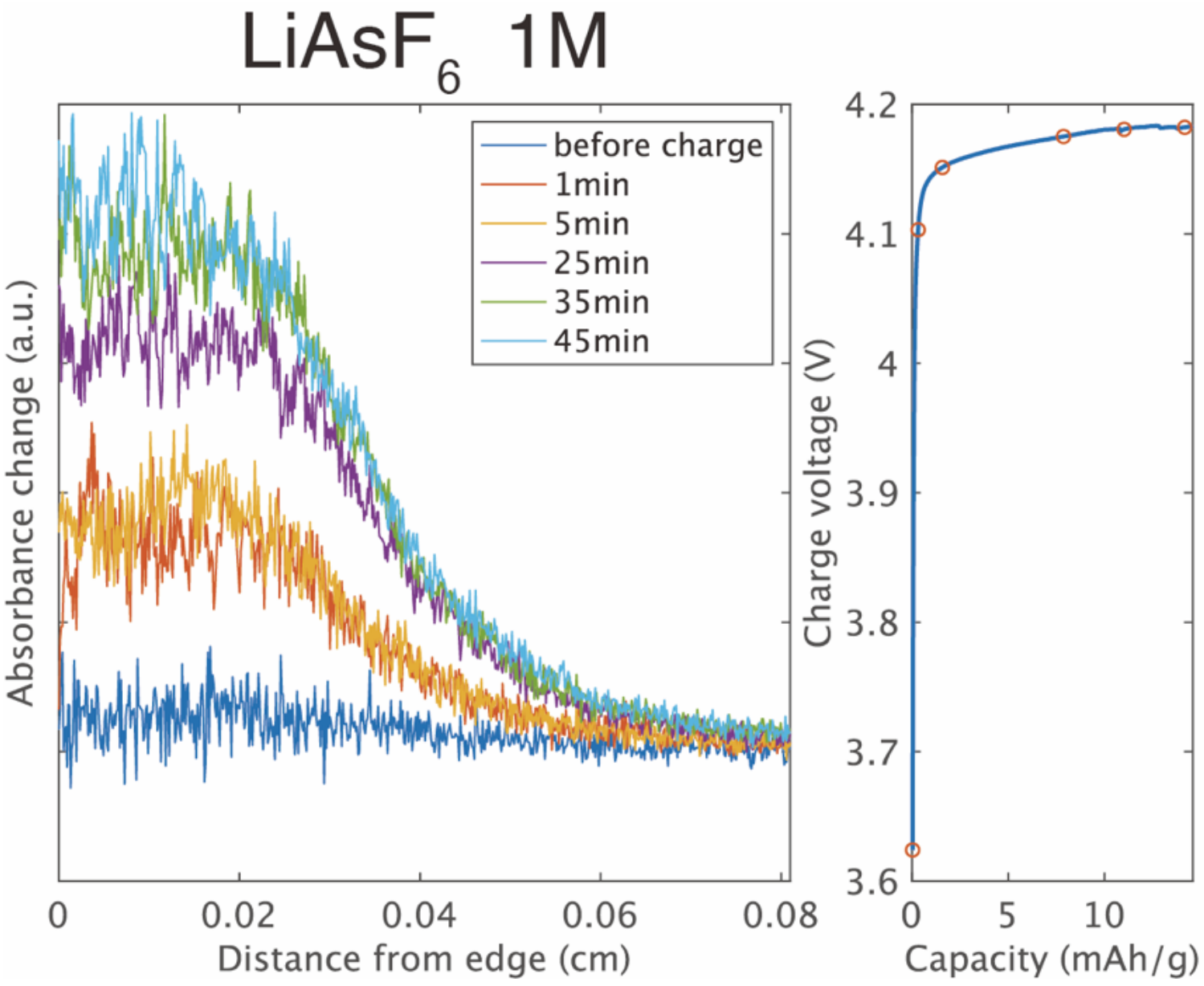

**Figure 7. Operando X-ray transmission measurement of spatial changes associated with electrolyte concentration.**

The left panel shows the temporal evolution of the X-ray absorbance change in a model electrode employing $LiAsF_6$ electrolyte. Because the X-ray attenuation of arsenic depends on the local electrolyte concentration, the absorbance change provides a spatially resolved measure of relative electrolyte-concentration changes during charging. The right panel shows the time evolution of the charging voltage and charge capacity for this sample electrode device.

# A physics-regularized neural network and Kirchhoff Markov random field framework for inferring latent electrochemical states from operando spectromicroscopy

Naoki Wada[1]†, Yuta Kimura[2], Masaichiro Mizumaki[3], Koji Amezawa[2], Ichiro Akai[4] and Toru Aonishi[1]†*

[1]Department of Complexity Science and Engineering, Graduate School of Frontier Sciences, The University of Tokyo, Kashiwa, Chiba 277-8561, Japan

[2]Institute of Multidisciplinary Research for Advanced Materials, Tohoku University, Sendai, Miyagi 980-8577, Japan

[3]Graduate School of Science and Technology, Kumamoto University, Kumamoto, Kumamoto 860-8555, Japan

[4]Institute of Industrial Nanomaterials, Kumamoto University, Kumamoto, Kumamoto 860-8555, Japan

†These authors contributed equally to this work.
*Correspondence: aonishi@edu.k.u-tokyo.ac.jp

# Supplementary Material

## S1 Physical parameters of the sample electrode

The model electrode has an in-plane area of $1\ \mathrm{cm} \times 1\ \mathrm{cm}$ and a thickness of $46\ \mu\mathrm{m}$. The theoretical specific capacity of LCO for full delithiation from $LiCoO_2$ to $CoO_2$ is 274 mAh g$^{-1}$ [1]. Given that the electrode contains 6.736 mg of LCO, its theoretical total capacity $Q_{\max}$ is $Q_{\max} = 274 \times 6.736 \times 10^{-3}$ mAh.

Let $L_z = 1$ cm denote the electrode length along the $z$-direction, $w = 1$ cm the electrode width, and $h = 46 \times 10^{-4}$ cm the electrode thickness. The spatial sampling interval is $\Delta x = 0.00065$ cm. The in-plane area assigned to one spatial subregion is therefore $A_{\mathrm{sub}} = w\Delta x = 6.5 \times 10^{-4}$ cm$^2$, whereas the electrode cross-sectional area perpendicular to the $z$-direction is $A_{\perp} = wh = 4.6 \times 10^{-3}$ cm$^2$.

Assuming that LCO is uniformly distributed over the electrode area, the capacity associated with each spatial subregion is proportional to its fractional area. Thus, letting $x_i$ denote the NLLS variable, the charge $q_i$ associated with the $i$-th subregion is given by

$$q_i = \frac{A_{\mathrm{sub}}}{A_{\mathrm{elec}}} x_i Q_{\max}. \qquad \text{(S1)}$$

Here, $A_{\mathrm{elec}} = wL_z = 1$ cm$^2$. Because $A_{\mathrm{sub}}/A_{\mathrm{elec}} = \Delta x / L_z$, Eq. (S1) can equivalently be written as

$$q_i = \frac{\Delta x}{L_z} x_i Q_{\max}. \qquad \text{(S2)}$$

Let $\Delta t$ (h) denote the measurement time interval. The LCO current density is defined as the rate of change of the charge associated with a subregion, normalized by the electrode cross-sectional area $A_{\perp}$. Accordingly, the scaling coefficient $S$ in Eq. (2) is

$$S = \frac{Q_{\max}}{A_{\perp}\Delta t} \frac{\Delta x}{L_z}. \qquad \text{(S3)}$$

The units of $S$ are mA cm$^{-2}$. Therefore, $I_{\mathrm{LCO},i}(t)$ in Eq. (2) represents the LCO current density associated with the $i$-th spatial subregion rather than a volumetric current density. In the present sign convention, $I_{\mathrm{LCO},i}(t)$ takes a negative value during charging.

Next, we estimate the exchange current density at the solid-liquid interface. The electrochemically active surface area per unit electrode volume, $a$ (cm$^2$ cm$^{-3}$), is evaluated following Ref. [2] as

$$a = \frac{3(1 - \varepsilon_{\mathrm{else}})}{R_s}. \qquad \text{(S4)}$$

Here, $R_s$ is the LCO particle radius and $\varepsilon_{\mathrm{else}}$ is the volume fraction of electrode components other than LCO. For the present electrode, $R_s = 6.0\ \mu\mathrm{m}$ and $\varepsilon_{\mathrm{else}} = 0.716$, yielding $a = 1419\ \mathrm{cm}^2\ \mathrm{cm}^{-3}$. The volume of one spatial subregion is $V_{\mathrm{sub}} = A_{\perp}\Delta x$, and the corresponding electrochemically active surface area is

$$A_{\mathrm{AM}} = aV_{\mathrm{sub}} = aA_{\perp}\Delta x. \qquad \text{(S5)}$$

The Newman-model expression gives the exchange current density per unit electrochemically active surface area as a function of the electrolyte salt concentration and the Li concentration in LCO. Multiplication by $A_{\mathrm{AM}}$ gives the total exchange current within one spatial subregion:

$$I_{0,\mathrm{sub}} = 0.1A_{\mathrm{AM}}kc_{\mathrm{salt}}^{0.5}(c_{\max} - c)^{0.5}c^{0.5}. \qquad \text{(S6)}$$

Here, $I_{0,\mathrm{sub}}$ is expressed in mA. The factor 0.1 converts the product of the exchange current density in A m$^{-2}$ and the electrochemically active surface area in cm$^2$ into a current in mA. $k$ is the reaction-rate constant [A m$^{5/2}$ mol$^{-3/2}$] including Faraday's constant, $c_{\mathrm{salt}}$ is the electrolyte salt concentration [mol m$^{-3}$], $c_{\max}$ is the maximum Li concentration in LCO [mol m$^{-3}$], and $c$ is the Li concentration in LCO [mol m$^{-3}$].

The exchange current density $i_0$ used in the Butler–Volmer relation in Eq. (5) is obtained by normalizing $I_{0,\mathrm{sub}}$ by the electrode cross-sectional area $A_{\perp}$:

$$i_0 = \frac{I_{0,\mathrm{sub}}}{A_{\perp}} = 0.1a\Delta x\, kc_{\mathrm{salt}}^{0.5}(c_{\max} - c)^{0.5}c^{0.5}. \qquad \text{(S7)}$$

Thus, $i_0$ has units of mA cm$^{-2}$. It represents the volumetric interfacial reaction integrated over the spatial interval $\Delta x$ and normalized by the electrode cross-sectional area. Accordingly, the interfacial current density $I_i(t)$, the LCO current density $I_{\mathrm{LCO},i}(t)$, and the liquid-phase current density $J_i(t)$ all have the same units of mA cm$^{-2}$, allowing the current-conservation relation to be written as $J_i(t) - J_{i-1}(t) + I_i(t) = 0$.

In the present study, the exchange current density was treated as a spatially and temporally constant effective parameter within each initial electrolyte-concentration condition. The parameters were set to $k = 5.0 \times 10^{-11}\ \mathrm{A\,m^{5/2}\,mol^{-3/2}}$, $c_{\max} = 52752\ \mathrm{mol\,m^{-3}}$, and $c = 40000\ \mathrm{mol\,m^{-3}}$. The value of $c$ was selected as a representative value within the range $32000$–$52752\ \mathrm{mol\,m^{-3}}$. The reaction-rate constant follows the value adopted in the NEDO RISING battery development project, while the remaining parameters were estimated from Ref. [1]. The electrolyte salt concentration $c_{\mathrm{salt}}$ was approximated by its initial value. Using $\Delta x = 0.00065\ \mathrm{cm}$, the exchange current densities used in Eq. (5) were $i_0 = 1.8040 \times 10^{-6}$, $3.2937 \times 10^{-6}$, and $4.6580 \times 10^{-6}\ \mathrm{mA\,cm^{-2}}$ for initial electrolyte concentrations of 0.3 M, 1 M, and 2 M, respectively.

The OCV curve $E(x)$ of LCO used in this study was obtained by regression of experimental data [3]:

$$E(x) = 2.162 + 0.471\tanh(2.766 - 5.703x) + 2.060\tanh(78.24 - 77.56x) + 0.142\tanh(4.123 - 8.023x). \qquad \text{(S8)}$$

Here, $E(x)$ is expressed in volts and $x$ is the NLLS variable. Because the solid–phase potential is taken as the reference in the equivalent–circuit model, the corresponding equilibrium liquid–side potential is defined as $V_{\mathrm{LCO},i}(t) = -E\left(y_{o,i}(t)\right)$.

The relationship between the effective ionic conductivity and electrolyte salt concentration in the positive electrode containing $LiPF_6$, shown in Fig. 6(a) of the main text, was calculated using the empirical relationship reported in Refs. [4,5]:

$$s_{\mathrm{eff}} = \varepsilon^{1.5}\left[-0.7222\left(\frac{c_{\mathrm{salt}}}{c_{\mathrm{ref}}}\right)^4 + 6.0577\left(\frac{c_{\mathrm{salt}}}{c_{\mathrm{ref}}}\right)^3 - 19.045\left(\frac{c_{\mathrm{salt}}}{c_{\mathrm{ref}}}\right)^2 + 22.614\left(\frac{c_{\mathrm{salt}}}{c_{\mathrm{ref}}}\right) + 0.311\right]. \qquad \text{(S9)}$$

Here, $s_{\mathrm{eff}}$ is expressed in mS cm$^{-1}$, $c_{\mathrm{salt}}$ is expressed in mol m$^{-3}$, and $c_{\mathrm{ref}} = 1000\ \mathrm{mol\,m^{-3}}$. Thus, $c_{\mathrm{salt}}/c_{\mathrm{ref}}$ is numerically equal to the salt concentration expressed in mol dm$^{-3}$ (M). The same definition and unit of $c_{\mathrm{salt}}$ are used in Eqs. (S7) and (S9).

The liquid–phase area–specific resistance $r_i(t)$ associated with the spatial interval $\Delta x$ is related to the effective ionic conductivity by $r_i(t) = \Delta x / s_{\mathrm{eff},i}(t)$. When $\Delta x$ is expressed in cm and $s_{\mathrm{eff}}$ in mS cm$^{-1}$, $r_i$ is expressed in k$\Omega$ cm$^2$. Consequently, $r_i(t)J_i(t)$ has units of volts when $J_i(t)$ is expressed in mA cm$^{-2}$.

## S2 Gibbs sampling

To sample from the high-dimensional probability density function defined by the Kirchhoff-MRF in Eq. (8), we employ an approximate Gibbs sampling algorithm based on locally Gaussian conditional distributions. Gibbs sampling is a type of Markov chain Monte Carlo (MCMC) method. At each update step, one variable is selected from the high-dimensional state variables, a sample is drawn from the conditional distribution of that variable, and the current state is updated with the sampled value. Repeating this procedure enables sampling from high-dimensional probability density functions or distributions for which direct sampling is difficult.

As described above, the probability density function treated in this study is independent at each time $t$. Therefore, Gibbs sampling is performed independently for each time. By completing the square, the conditional probability density functions for each state variable $I_i$, $J_i$, $V_i$, and $r_i$ can be derived as Eqs. (S10)–(S21):

$$P(I_i \mid \{I_{/i}\}, \{J_i\}, \{V_i\}, \{r_i\}, \{I_{\mathrm{LCO},i}(t)\}, \{V_{\mathrm{LCO},i}(t)\}) \propto \exp\left(-\kappa_{I_i}\left(I_i - \frac{\mu_{I_i}}{\kappa_{I_i}}\right)^2\right), \tag{S10}$$

$$\kappa_{I_i} = \beta + \delta + \eta\left[\frac{1}{BV'(V_i - V_{\mathrm{LCO},i}(t))}\right]^2, \tag{S11}$$

$$\mu_{I_i} = \beta I_{\mathrm{LCO},i}(t) + \delta(J_{i-1} - J_i) + \eta\frac{BV(V_i - V_{\mathrm{LCO},i}(t))}{BV'(V_i - V_{\mathrm{LCO},i}(t))^2}, \tag{S12}$$

$$P(J_i \mid \{I_i\}, \{J_{/i}\}, \{V_i\}, \{r_i\}, \{I_{\mathrm{LCO},i}(t)\}, \{V_{\mathrm{LCO},i}(t)\}) \propto \exp\left(-\kappa_{J_i}\left(J_i - \frac{\mu_{J_i}}{\kappa_{J_i}}\right)^2\right), \tag{S13}$$

$$\kappa_{J_i} = 2\delta + \gamma r_i^2, \tag{S14}$$

$$\mu_{J_i} = \delta(J_{i+1} + J_{i-1} + I_{i+1} - I_i) - \gamma r_i(V_{i+1} - V_i), \tag{S15}$$

$$P(V_i \mid \{I_i\}, \{J_i\}, \{V_{/i}\}, \{r_i\}, \{I_{\mathrm{LCO},i}(t)\}, \{V_{\mathrm{LCO},i}(t)\}) \propto \exp\left(-\kappa_{V_i}\left(V_i - \frac{\mu_{V_i}}{\kappa_{V_i}}\right)^2\right), \tag{S16}$$

$$\kappa_{V_i} = 2\gamma + \eta, \tag{S17}$$

$$\mu_{V_i} = \gamma(V_{i+1} + V_{i-1} + r_i J_i - r_{i-1}J_{i-1}) + \eta\left(BV^{-1}(I_i) + V_{\mathrm{LCO},i}(t)\right), \tag{S18}$$

$$P(r_i \mid \{I_i\}, \{J_i\}, \{V_i\}, \{r_{/i}\}, \{I_{\mathrm{LCO},i}(t)\}, \{V_{\mathrm{LCO},i}(t)\}) \propto \exp\left(-\kappa_{r_i}\left(r_i - \frac{\mu_{r_i}}{\kappa_{r_i}}\right)^2\right), \tag{S19}$$

$$\kappa_{r_i} = \gamma J_i^2 + 2\phi + \psi, \tag{S20}$$

$$\mu_{r_i} = -\gamma J_i(V_{i+1} - V_i) + \phi(r_{i+1} + r_{i-1}) + \psi\bar{r}. \tag{S21}$$

Here, $\{X_{/i}\}$ denotes the set of variables $X_k(k = 1, \ldots, i-1, i+1, \ldots, N)$ excluding the $i$-th variable $X_i$. As mentioned above, because Eq. (10) includes the Butler-Volmer (BV) function that describes the nonlinearity at the solid-liquid interface, the high-dimensional probability density function defined by Eq. (8) is not a multivariate normal distribution. Therefore, in deriving Eqs. (S10)-(S12), we perform a first-order perturbation expansion of the BV function and approximate the conditional distributions as normal distributions. In addition, sampling at both ends of the electrode is performed based on the conditional probability density functions in which the boundary-condition relations in Eq. (6) are substituted.

Based on these conditional probability density functions, we construct the Gibbs sampling algorithm shown in Algorithm S1 and approximate the marginal posterior distributions.

**Algorithm S1 Approximate Gibbs sampling for the Kirchhoff-MRF**

| | |
|---|---|
| **Input** | $\{V_{LCO,i}(t)\}$, $\{I_{LCO,i}(t)\}$, hyperparameters $\beta, \delta, \gamma, \eta, \phi,$ and $\psi$; reference area-specific resistance $\bar{r}$; total number of Gibbs sweeps $\mathrm{M_{total}}$; burn-in length $\mathrm{M_{burn}}$; and thinning interval $\mathrm{M_{thin}}$. |
| **Output** | Approximate marginal posterior samples and posterior-mean estimates of $\{I_i\}, \{J_i\}, \{V_i\},$ and $\{r_i\}$. |

**1.** Initialize $\{I_i\}_{i=1,\cdots,\mathrm{N}}, \{J_i\}_{i=1,\cdots,\mathrm{N}}, \{V_i\}_{i=1,\cdots,\mathrm{N}},$ and $\{r_i\}_{i=1,\cdots,\mathrm{N}}$.

**2.** Set the retained-sample set $\mathrm{S} = \emptyset$.

**3.** for $m = 1, \cdots, \mathrm{M_{total}}$ do

**3.1.** for $i = 1, \cdots,$ N do

Draw the interfacial current density from its conditional distribution:

$$I_i^{(\mathrm{new})} \sim \mathcal{N}\left(\frac{\mu_{I_i}}{\kappa_{I_i}}, \frac{1}{2\kappa_{I_i}}\right)$$

and update $I_i \leftarrow I_i^{(\mathrm{new})}$.

Draw the liquid-phase current density from its conditional distribution:

$$J_i^{(\mathrm{new})} \sim \mathcal{N}\left(\frac{\mu_{J_i}}{\kappa_{J_i}}, \frac{1}{2\kappa_{J_i}}\right)$$

and update $J_i \leftarrow J_i^{(\mathrm{new})}$.

Draw the liquid-phase potential from its conditional distribution:

$$V_i^{(\mathrm{new})} \sim \mathcal{N}\left(\frac{\mu_{V_i}}{\kappa_{V_i}}, \frac{1}{2\kappa_{V_i}}\right)$$

and update $V_i \leftarrow V_i^{(\mathrm{new})}$.

Draw the liquid-phase area-specific resistance from its conditional distribution:

$$r_i^{(\mathrm{new})} \sim \mathcal{N}\left(\frac{\mu_{r_i}}{\kappa_{r_i}}, \frac{1}{2\kappa_{r_i}}\right)$$

and update $r_i \leftarrow \max\left(r_i^{(new)}, \Delta x / s_{\mathrm{bulk,max}}\right)$.

At the two electrode boundaries, use the corresponding conditional distributions obtained by substituting the boundary conditions in Eq. (6).

**3.2.** end for

**3.3.** After completing one full Gibbs sweep, if

$$m > \mathrm{M_{burn}} \text{ and } (m - \mathrm{M_{burn}}) \bmod \mathrm{M_{thin}} = 0,$$

retain the current state:

$$S \leftarrow S \cup \left\{\left\{I_i^{(m)}\right\}, \left\{J_i^{(m)}\right\}, \left\{V_i^{(m)}\right\}, \left\{r_i^{(m)}\right\}\right\}.$$

**4.** end for

**5.** Using the L retained samples, calculate the posterior-mean estimates:

$$\hat{I}_i = \frac{1}{\mathrm{L}}\sum_{l=1}^{\mathrm{L}} I_i^{(l)}, \qquad \hat{J}_i = \frac{1}{\mathrm{L}}\sum_{l=1}^{\mathrm{L}} J_i^{(l)},$$

$$\hat{V}_i = \frac{1}{\mathrm{L}}\sum_{l=1}^{\mathrm{L}} V_i^{(l)}, \qquad \hat{r}_i = \frac{1}{\mathrm{L}}\sum_{l=1}^{\mathrm{L}} r_i^{(l)}.$$

**6.** Return the retained samples and the posterior-mean estimates $\{\hat{I}_i\}, \{\hat{J}_i\}, \{\hat{V}_i\}$, and $\{\hat{r}_i\}$.

**Note.** The sampler was run for 1,500,001 iterations. The first 500,000 iterations were discarded as burn-in, and every 100th sample was retained. Here, $\Delta x = 0.00065$ cm is the spatial sampling interval, and $s_{\text{bulk,max}} = 9.2\ \mathrm{mS\,cm^{-1}}$ is the maximum conductivity measured for the bulk liquid electrolyte. The constraint $r_i \geq \Delta x / s_{\text{bulk,max}}$ is used only to prevent unrealistically small resistance values during sampling. The physical admissibility of the reconstructed effective conductivity within the electrode is assessed separately using the criterion described in Section S3.

# S3 Hyperparameter setting

Supplementary Table S1 lists the values of the regularization parameters introduced in Eq. (4) for the three-layer NN. For the regularization parameter $A$ associated with current conservation and the regularization parameter $C$ that constrains the admissible NLLS range, we set sufficiently large values within a range that does not cause numerical divergence, because these constraints must be satisfied physically and violations should not occur during training. This strongly suppresses violations of current conservation and constrains the estimated NLLS to remain within a physically reasonable range.

In contrast, the regularization parameter $B$ controlling the spatial continuity of NLLS was selected empirically by considering the trade-off between excessive smoothing and noise amplification. Similarly, the number of hidden-layer units $K$ was selected empirically so that the stability and reproducibility of the estimated spatiotemporal NLLS map are ensured.

Supplementary Table S2 lists the hyperparameter values used in the Kirchhoff-MRF of Eq. (7). The Kirchhoff-MRF is a hierarchical Bayesian model as shown in Fig. 4(c), and in principle, it is possible to sample the hyperparameters simultaneously. However, at present, it is difficult to obtain numerically stable convergence when estimating the hyperparameters together with the latent variables. Therefore, in this study, the hyperparameters are fixed to predetermined values.

For $\delta$, the hyperparameter corresponding to the term enforcing Kirchhoff's first law, we set a sufficiently large value relative to the likelihood term within a range that does not make the estimation process numerically unstable, because this physical law must be satisfied as a constraint. This suppresses substantial deviations of the estimated current distribution from the physical law.

In addition, for the hyperparameter $\psi$ associated with the regularization term that constrains the liquid-phase ohmic area-specific resistance $r_i(t)$ to fluctuate around the reference area-specific resistance $\bar{r}$, which is determined from the initial electrolyte concentration, we set $\psi = 1/\bar{r}^2$ so that the magnitude of the fluctuations is suppressed to a level comparable to $\bar{r}$. Furthermore, for the hyperparameter $\phi$ corresponding to the regularization term that enforces spatial continuity of the liquid-phase area-specific resistance, we set $\phi = 0.1\beta$, taking into account both the balance with the likelihood term associated with the observation error of the current and the difference in scale among the variables.

On the other hand, it was confirmed that the hyperparameter $\gamma$ for the term including the ohmic voltage loss in Li-ion transport along the $z$-direction, as well as the hyperparameter $\eta$ for the term corresponding to the Butler-Volmer equation describing the solid-liquid interfacial reaction, exert pronounced effects on the scale and physical plausibility of the estimated liquid-phase conductivity and potential distributions. Therefore, in this study, for each of $\gamma$ and $\eta$, we prepared ten candidate values in the ranges of $0.01$–$10$ times $\beta$ and $10^{-6}$–$10^{-3}$ times $\beta$, respectively, and determined $\gamma$ and $\eta$ by performing a two-dimensional grid search on the $\gamma$–$\eta$ plane.

Supplementary Figure S1 shows the results of the two-dimensional grid search on the $\gamma$–$\eta$ plane conducted under the initial electrolyte concentration of 0.3 M.

Supplementary Figure S1(a) shows the $\gamma$–$\eta$ dependence of the liquid-phase potential gradient near the electrode edge at the end of charging. Physically, the potential gradient from the electrode edge toward the interior is required to be positive; therefore, the region where the potential gradient is positive represents a valid parameter range. When $\gamma$ is small, the potential gradient tends to become negative.

Supplementary Figure S1(b) shows the dependence of the maximum estimated liquid-phase conductivity, evaluated over all times and spatial positions, on $\gamma$ and $\eta$. As $\eta$ increases, this maximum tends to exceed 3.4 mS $cm^{-1}$, the upper limit adopted for the effective conductivity within the electrode based on the empirical conductivity–concentration relationship in Eq. (S9). Parameter combinations exceeding this limit were therefore regarded as physically inadmissible. This criterion is distinct from the bulk-electrolyte conductivity bound used solely as a numerical safeguard during sampling in Algorithm S1.

Supplementary Figure S1(c) shows the $\gamma$–$\eta$ dependence of the difference between the estimated liquid-phase conductivity at the electrode edge and that at the electrode center at the beginning of charging. This sign condition was specified a priori on the basis of the electrochemical response expected under the 0.3 M condition, rather than being inferred from the reconstructed conductivity distribution. At the beginning of charging, delithiation occurs preferentially near the electrode edge, increasing the local electrolyte salt concentration relative to that in the electrode interior. Because 0.3 M lies on the increasing branch of the conductivity–concentration curve, for which $ds/dc_{\mathrm{salt}} > 0$, the resulting conductivity near the electrode edge is required to be higher than that at the electrode center within the assumptions of the present model. Parameter combinations that produced the opposite sign were therefore regarded as physically inadmissible.

The physically admissible parameter region was defined as the intersection of the following three independently specified conditions: (i) the liquid-phase potential gradient from the electrode edge toward the interior is positive; (ii) the maximum liquid-phase conductivity remains within the range permitted by the empirical conductivity–concentration relationship; and (iii) under the 0.3 M condition, the liquid-phase conductivity at the electrode edge is higher than that at the electrode center at the beginning of charging. Condition (iii) constrains only the sign of the initial edge-to-center conductivity difference and does not prescribe its magnitude, subsequent temporal evolution, or spatial propagation. The parameter pair indicated by the black circle was selected from the region satisfying these physical admissibility conditions and was subsequently applied to all three initial electrolyte-concentration conditions.

For initial electrolyte concentrations of 1 M and 2 M, the estimated conductivity at the electrode edge tended to be consistently lower than that at the electrode center regardless of $\gamma$. In addition, the potential gradient was always positive regardless of $\gamma$, satisfying physical consistency. In contrast, the spatial variations in the estimated liquid-phase conductivity were strongly dependent on $\eta$: when $\eta$ was set too large, the estimated conductivity exhibited temporally and spatially unstable oscillations, tending to produce physically implausible behavior. Based on these results, we confirmed that using the same hyperparameter setting selected under the 0.3 M condition yields stable and physically reasonable estimates for both 1 M and 2 M. Therefore, the same hyperparameters were employed for all initial electrolyte concentration conditions in this study.

**S4. Sensitivity analysis of the Kirchhoff-MRF hyperparameters**

To assess whether the principal reconstructed features depended strongly on the selected hyperparameters, we performed a local sensitivity analysis around the adopted parameter pair $(\gamma^*, \eta^*)$. Each parameter was independently perturbed to the immediately adjacent values in the original grid while the other parameter was held fixed. Accordingly, five parameter pairs were examined: $(\gamma^*, \eta^*)$, $(\gamma_-, \eta^*)$, $(\gamma_+, \eta^*)$, $(\gamma^*, \eta_-)$, and $(\gamma^*, \eta_+)$. Each parameter value is shown in Table S3.

For each parameter pair, we evaluated the edge-to-center conductivity contrast,

$$\Delta s(t) = \bar{s}_{\mathrm{edge}}(t) - \bar{s}_{\mathrm{center}}(t),$$

together with the residuals of the current-conservation and liquid-phase Ohmic relations defined as the following:

$$R_{KCL} = \sqrt{\frac{\frac{1}{NT}\sum_{t=1}^{T}\sum_{i=2}^{N}\left[J_i(t) - J_{i-1}(t) + I_i(t)\right]^2}{\frac{1}{NT}\sum_{t=1}^{T}\sum_{i=1}^{N} J_i(t)^2}},$$

$$R_{ohm} = \sqrt{\frac{\frac{1}{NT}\sum_{t=1}^{T}\sum_{i=1}^{N-1}\left[V_{i+1}(t) - V_i(t) + r_i(t)J_i(t)\right]^2}{\frac{1}{NT}\sum_{t=1}^{T}\sum_{i=1}^{N}V_i(t)^2}}.$$

Although the perturbations produced quantitative variations in the reconstructed conductivity fields, the principal contrast between the 0.3 M condition and the 1 M and 2 M conditions was preserved across the examined parameter sets, as shown in Fig. S2(a). Moreover, the normalized residuals of the current-conservation and liquid-phase Ohmic relations remained at comparable levels across the examined parameter sets, as shown in Fig. S2(b). These results indicate that the qualitative conclusions are not specific to the single adopted parameter pair.

Table S1. Regularization parameters of the three-layer NN

| Parameter | Value |
|---|---|
| $\boldsymbol{A}$ | 30 |
| $\boldsymbol{B}$ | 30 |
| $\boldsymbol{C}$ | 30 |
| $\boldsymbol{K}$ | 2 |

Table S2. Hyperparameters of the Kirchhoff-MRF

| Parameter | Value |
|---|---|
| $\boldsymbol{\beta}$ | $10^6$ |
| $\boldsymbol{\delta}$ | $10\beta$ |
| $\boldsymbol{\gamma}$ | $0.2\beta$ |
| $\boldsymbol{\eta}$ | $5 \times 10^{-5}\beta$ |
| $\boldsymbol{\phi}$ | $0.1\beta$ |
| $\boldsymbol{\psi}$ | $1/\bar{r}^2$ |

Table S3. Parameters of the sensitivity analysis

| Parameter | Value |
|---|---|
| $\boldsymbol{\beta}$ | $10^6$ |
| $\boldsymbol{\gamma}^*$ | $0.2\beta$ |
| $\boldsymbol{\gamma}_-$ | $0.1\beta$ |
| $\boldsymbol{\gamma}_+$ | $0.3\beta$ |
| $\boldsymbol{\eta}^*$ | $5 \times 10^{-5}\beta$ |
| $\boldsymbol{\eta}_-$ | $2 \times 10^{-5}\beta$ |
| $\boldsymbol{\eta}_+$ | $7 \times 10^{-5}\beta$ |

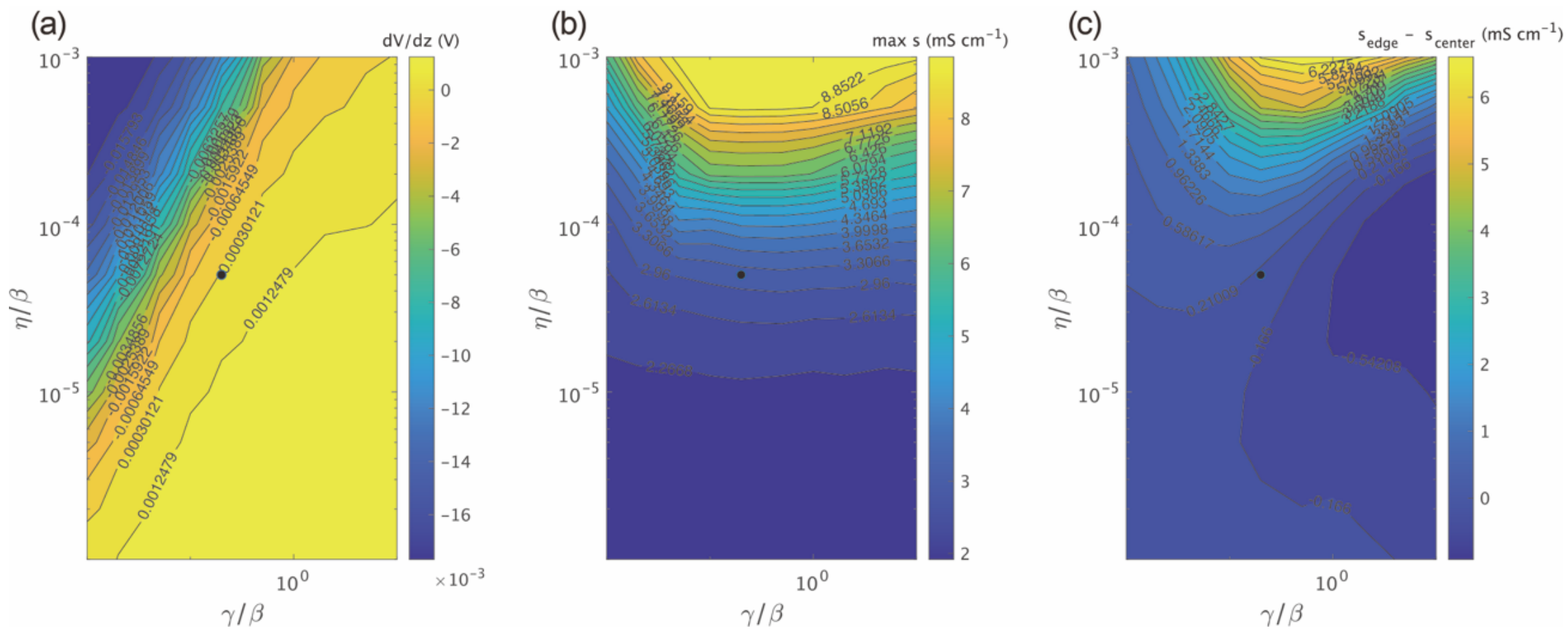

**Figure S1. Results of the two-dimensional grid search on the $\gamma-\eta$ plane.**

Dependence of the estimation results on $\gamma$ and $\eta$ under the initial electrolyte concentration condition of 0.3 M. **(a)** $\gamma-\eta$ dependence of the liquid-phase potential gradient near the electrode edge at the end of charging. **(b)** $\gamma-\eta$ dependence of the maximum value of the estimated liquid-phase conductivity over the entire spatiotemporal domain. **(c)** $\gamma-\eta$ dependence of the edge-to-center conductivity difference, $s_{\text{edge}} - s_{\text{center}}$, at the beginning of charging. The black circle indicates the values of $\gamma$ and $\eta$ adopted in this study.

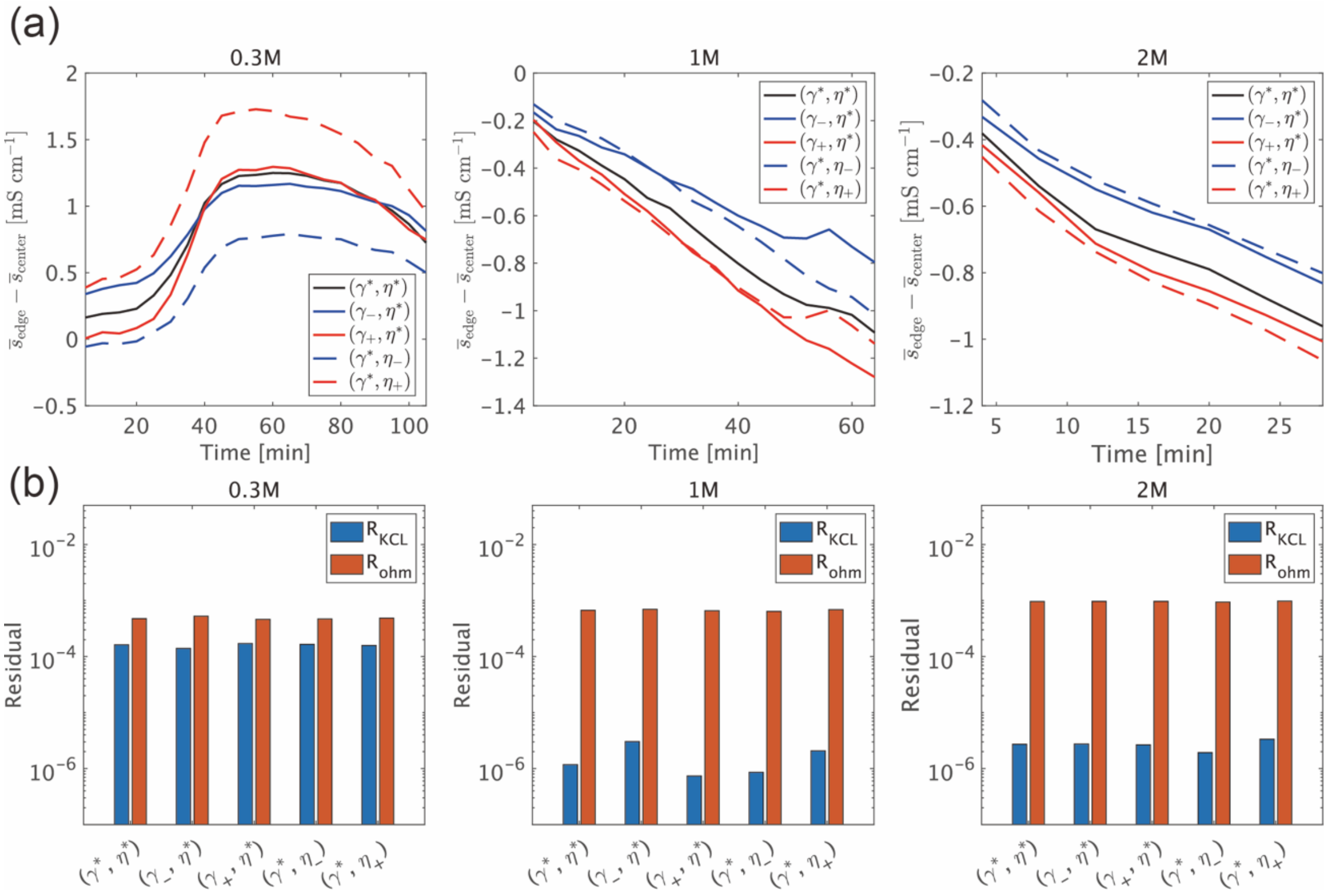

**Figure S2. Local sensitivity of the reconstructed electrolyte conductivity to the Kirchhoff–MRF hyperparameters.**

The hyperparameters $\gamma$ and $\eta$ were independently varied around the adopted parameter pair $(\gamma^*, \eta^*)$. **(a)** The curves show the conductivity contrast $\Delta s = \bar{s}_{\text{edge}} - \bar{s}_{\text{center}}$ for the 0.3, 1, and 2 M conditions. The adopted parameter pair is indicated by the thick black line, whereas the four neighboring parameter pairs are indicated by colored solid or dashed lines. **(b)** The bars show the residuals of the current-conservation and liquid-phase Ohmic relations, $R_{KCL}$ and $R_{ohm}$ for the 0.3, 1, and 2 M conditions.